\documentclass{JHEP3}
\usepackage{amsmath}
\usepackage{amssymb}

\usepackage{epsfig}

\def\beq{\begin{equation}}
\def\eeq{\end{equation}}
\def\beqa{\begin{eqnarray}}
\def\eeqa{\end{eqnarray}}
\def\bet{\begin{tabular}}
\def\eet{\end{tabular}}

\newcommand{\sect}[1]{\setcounter{equation}{0}\section{#1}}

\newcommand{\msbar}{$\overline{\mbox{MS}}$ }

\def \lsim
{\raisebox{-3pt}{$\>\stackrel{<}{\scriptstyle\sim}\>$}}
\def \gsim
{\raisebox{-3pt}{$\>\stackrel{>}{\scriptstyle\sim}\>$}}

\title{Charm-quark fragmentation\\ with an effective coupling 
constant}

\vspace{5mm}
\author{Gennaro Corcella\\
Dipartimento di Fisica, Universit\`a di Roma `La Sapienza'\\ 
P.le A.~Moro~2, I-00185 Roma, Italy.\\
E-mail: \email{Gennaro.Corcella@roma1.infn.it}}

\author{Giancarlo Ferrera\\ Departament d'Estructura i Constituents de la
Mat\`eria,\\
Universitat de Barcelona, \\
Diagonal 647, E-08028 Barcelona, Spain \\
E-mail: \email{ferrera@ecm.ub.es}}

\abstract{
We use a recently proposed non-perturbative model, based on an 
effective strong coupling constant and free from tunable parameters,
to study $c$-flavoured hadron
production in $e^+e^-$ annihilation. Charm-quark production is described
in the framework of perturbative fragmentation functions, 
with NLO coefficient
functions, NLL non-singlet DGLAP evolution and NNLL large-$x$ resummation.
We model hadronization effects by means of the effective coupling 
constant in the
NNLO approximation and compare our results with experimental data
taken at the $Z^0$ pole and at the $\Upsilon(4S)$ resonance.
We find that, within the experimental and 
theoretical uncertainties, our model is able to 
give a reasonable description of 
${D^*}^+$-meson spectra from ALEPH 
for $x<1-\Lambda/m_c$. More serious discrepancies
are instead present when comparing with $D$ and $D^*$
data from BELLE and CLEO in $x$-space. 
Within the errors, our model is nonetheless capable of reproducing the
first ten Mellin moments of all considered data sets.
However, the fairly large theoretical 
uncertainties call for a full NNLO/NNLL analysis.}

\preprint{ROME1/1453/07\\UB-ECM-PF-07-16
\\December 2007}


\begin{document}

\section{Introduction}
\label{intro}

The hadronization of partons into hadrons, for the time being, cannot
be calculated from first principles, but it is usually described in
terms of phenomenological models, such as the Kartvelishvili \cite{kart}
or Peterson \cite{pet} non-perturbative fragmentation functions,
containing few parameters which need to be tuned to experimental data.
It was recently proposed \cite{afr,bfrag}, however, a 
non-perturbative
model, based on the work in refs.~\cite{shirkov,stefanis}, including
power corrections via an effective strong coupling constant, which does not
exhibit the Landau pole any longer and includes absorptive effects due
to gluon branching.
The interesting feature of such a model is that it does not contain
any extra free parameter to be fitted to the data, besides the
ones entering in the parton-level calculation. 
In \cite{afr}, such a model was used in the framework 
of $B$-meson decays and 
it was found good agreement with the data on the photon spectrum
and on the hadron-mass distribution in radiative and
semileptonic decays, respectively.
In \cite{bfrag} the effective coupling was employed in the framework
of bottom-quark fragmentation and, within the theoretical uncertainties,
a reasonable fit of LEP and
SLD data on $B$-hadron spectra was obtained in both $x$ and moment spaces.

Although the results in \cite{afr,bfrag} are encouraging, it is nonetheless
mandatory to consider more data and observables to validate the 
effective-coupling model.
In this paper, we consider charm-quark fragmentation
in $e^+e^-$ processes and investigate
how our non-perturbative model fares against $D$-meson 
data from LEP and $B$-factories. In fact,
charm production involves pretty different scales with respect to 
$b$-quark fragmentation, and therefore the comparison with
$D$-hadron spectra should help to shed light on our model.
Considering charm production at 
the $Z^0$ pole and at the $\Upsilon(4S)$ resonance, furthermore,
is also interesting to understand how our model behaves when the 
process hard scale changes.
Perturbative charm production will be described in the 
framework of perturbative fragmentation functions \cite{mele}, 
using the same approximations carried out in
\cite{bfrag}, and the effective coupling constant will be our only source of
non-perturbative power corrections.

The plan of the present paper is the following.
In section 2 we shall review the main points of the parton-level
computation, based on the perturbative fragmentation formalism, and
including large-$x$ resummation in both coefficient function and
initial condition of the perturbative fragmentation function.
In section 3 we shall discuss the effective coupling constant and the
inclusion of non-perturbative corrections
to charm-quark fragmentation.
In section 4 we shall compare the results with charmed-meson spectra
from LEP and
$B$-factories in $x$-space, whereas we present our analysis in
Mellin moment space in section 5.
We shall finally summarize our main results in section 6.

\sect{Charm-quark production}
\label{quark}

In this section we shall 
discuss our calculation for charm-quark production.
For the sake of consistency, and given the tight relation between
perturbative calculation and non-perturbative corrections,
our computation will be carried out along the lines of
ref.~\cite{bfrag}.
Therefore, we shall just point out
the main issues involved in the calculation
and refer to \cite{bfrag} for further details.

\subsection{Perturbative fragmentation functions}
\label{sub1}

We consider $c\bar c$-pair production in $e^+e^-$ annihilation
at next-to-leading order (NLO) in the strong coupling constant $\alpha_S$:
\begin{equation}
e^+e^-\to P(Q) \to c(p_c) \bar c(p_{\bar c}) \left( g(p_g)\right)
\end{equation}
and define the charm-quark energy fraction:
\begin{equation}
x={{2p_c\cdot Q}\over{Q^2}}.
\end{equation}
In the following, we shall consider charm production at LEP,
where $P$ is a $Z^0$ boson and $Q=m_Z$, as well as
$c$-quark fragmentation at the $\Upsilon(4S)$ 
resonance, i.e. $Q=m_{\Upsilon(4S)}$
and the $c\bar c$ pair coming from the decay of a virtual
photon ($P=\gamma^*$).
In principle, in $e^+e^-$ annihilation 
charm quarks can also come from other processes, such
as the decay of bottomed hadrons 
produced via $Z^0 (\Upsilon(4S))\to b\bar b$.
However, as we shall discuss in detail in section 4, 
our analysis will only deal with direct $c\bar c$ production.

The perturbative fragmentation approach \cite{mele}, 
up to power corrections, 
factorizes the 
energy distribution of a heavy quark, the charm quark in our case, 
as the 
convolution of a coefficient function, associated with the emission off a 
massless parton, and a perturbative fragmentation function, expressing the
transition of the light parton into a heavy quark. 
This way, the $c$-quark spectrum reads:
\begin{eqnarray}
{1\over {\sigma}} {{d\sigma}\over{dx}} (x,Q,m_c) &=&
\sum_i\int_{x}^1
{{{dz}\over z}\left[{1\over{\sigma}}
{{d\hat\sigma_i}\over {dz}}(z,Q,\mu_R,\mu_F)
\right]^{\overline{\mathrm{MS}}}
D_i^{\overline{\mathrm{MS}}}\left({x\over z},\mu_F,m_c \right)} \nonumber \\
&+& {\cal O}\left((m_c/Q)^p\right) .
\label{pff}
\end{eqnarray}
In eq.~(\ref{pff}), $p\geq 1$, $d\hat\sigma_i /dz$ is the differential 
cross section for the 
production of a massless parton $i$
after subtracting the collinear singularity in the \msbar factorization
scheme; $\mu_R$ and $\mu_F$ are the renormalization and factorization
scales; $\sigma$ is the NLO $e^+e^-\to q\bar q(g)$ cross section.
Hereafter, we shall neglect charm production via
$g\to c\bar c$ splitting. In fact, we can anticipate that,
when comparing with data, secondary $c\bar c$ 
production will be either subtracted off the sample which we shall analyse
or negligible at the centre-of-mass considered energy.
This implies that
$i=c$ in eq.~(\ref{pff}) 
and $D_c^{\overline{\mathrm{MS}}}$ is the perturbative fragmentation
function expressing the fragmentation of a massless $c$ into a massive $c$.
The NLO \msbar coefficient function for $e^+e^-\to q\bar q$ processes
can be found in 
\cite{marti}.

The perturbative fragmentation function follows the 
DGLAP evolution equations
\cite{dgl,ap}; its value at a any scale $\mu_F$ can be obtained once
an initial condition at $\mu_{0F}$ 
is given. In \cite{mele} the initial condition
$D_c^{\rm ini}(x,\mu_{0F},m_c)$ was calculated in the NLO approximation
and its process-independence 
was established on more general grounds in \cite{cc}.
It is given by:
\begin{equation}
D_c^{\rm ini}(x,\alpha_S(\mu_{0R}^2),\mu_{0F}^2,m_c^2)=\delta(1-x)+
{{\alpha_S(\mu_{0R}^2)C_F}\over{2\pi}}
\left[{{1+x^2}\over{1-x}}\left(\ln {{\mu_{0F}^2}\over{m_c^2}}-
2\ln (1-x)-1\right)\right]_+.
\label{dbb}
\end{equation}

As discussed in \cite{mele}, solving the DGLAP equations for an evolution
from $\mu_{0F}$ to $\mu_F$, with a NLO kernel, allows one to resum leading (LL)
$\alpha_S^n\ln^n(\mu_F^2/\mu_{0F}^2)$ and next-to-leading (NLL) 
$\alpha_S^n \ln^{n-1}(\mu_F^2/\mu_{0F}^2)$ logarithms.
Setting $\mu_{0F}\simeq m_c$ and $\mu_F\simeq Q$, one resums the large
$\ln(Q^2/m_c^2)$ appearing in the massive NLO spectrum \cite{mele}.
The resummation of such mass logarithms is usually called
collinear resummation.
For the sake of working in the same perturbative framework as in \cite{bfrag},
in the following we shall consider NLO coefficient functions and initial 
condition,
along with NLL non-singlet DGLAP evolution. However, one could go beyond
such a level of accuracy and include 
NNLO corrections to the coefficient function \cite{neerven,moch}, initial
condition \cite{alex1} and to the non-singlet splitting functions 
\cite{alex2} entering in the kernel of the DGLAP equations.
The gluon-initiated contribution to the initial condition, necessary
to possibly extend the analysis to the singlet sector,  was calculated
in \cite{mele} and \cite{alex3} to NLO and NNLO, respectively.

\subsection{Large-$x$ resummation}
\label{sub2}

Both coefficient function \cite{mele} and initial condition (\ref{dbb})
contain terms, $\sim 1/(1-x)_+$ and $\sim[\ln(1-x)/(1-x)]_+$, 
enhanced
when $x$ approaches 1, which corresponds to soft- or collinear-gluon radiation.
One needs to resum such contributions to all orders to improve the
perturbative prediction (threshold resummation).
As in \cite{bfrag}, we shall implement threshold resummation, which
is process-dependent in the coefficient function and process-independent in 
the initial condition \cite{cc}, in the next-to-next-to-leading
logarithmic (NNLL) approximation, following the
general method of \cite{sterman,ct}.
Large-$x$ resummation is typically performed in Mellin moment-space,
where the Mellin transform of the differential cross section reads:
\beq
\sigma_N=\int_0^1dx\ x^{N-1}{1\over\sigma}{{d\sigma}\over{dx}}\ .\eeq
In $N$-space, the enhanced contributions $\sim \alpha_S/(1-x)_+$ and 
$\alpha_S[\ln(1-x)/(1-x)]_+$ correspond to single $(\sim \alpha_S\ln N)$
and double ($\sim \alpha_S\ln^2 N$) logarithms of the Mellin variable $N$.
The resummed coefficient function is given by the following
generalized exponential function \cite{cc}:
\begin{equation}\label{rescf}
\Delta_N^{(C)}\left[\alpha_S(\mu_R^2), \mu_R^2, \mu_F^2, Q^2\right] \, = \, 
\exp\left\{
G_N^{(C)}\left[ \alpha_S(\mu_R^2), \mu_R^2, \mu_F^2, Q^2 \right]
\right\},
\end{equation}
where 
\begin{eqnarray}
\label{GNC}
G_N^{(C)}\left[ \alpha_S(\mu_R^2), \mu_R^2, \mu_F^2, Q^2 \right] \, &=& \, 
\int_0^1 {dz {{z^{N-1}-1}\over{1-z}}}
\left\{\int_{\mu_F^2}^{Q^2 (1-z)} {{dk^2}\over {k^2}}
A\left[\alpha_S(k^2)\right]\right.\nonumber\\ &+&\left.
B\left[\alpha_S\left(Q^2(1-z)\right)\right]\right\}.
\end{eqnarray}
The exponent
$G^{(C)}_N\left[ \alpha_S(\mu_R^2), \mu_R^2, \mu_{F}^2, Q^2 \right]$ 
resums the large logarithms of the Mellin variable;
in the NNLL approximation, one keeps in the exponent terms 
$\sim \alpha_S^n\ln^{n+1}N$ (LL),
$\sim \alpha_S^n\ln^nN$ (NLL) and $\sim \alpha_S^n\ln^{n-1}N$ (NNLL). 
As in \cite{ct}, the integration variables are 
$z \, = \, 1 - x_g$, $x_g$ being the gluon energy fraction, and 
$k^2 \, = \, (p_c+p_g)^2(1-z)$. In soft approximation, $z\simeq x$; for
small-angle radiation $k^2 \, \simeq \, k_{\perp}^2$, 
the gluon transverse momentum with respect to the $c$.

In (\ref{GNC}), 
function $A(\alpha_S)$ resums soft and collinear radiation, while
$B(\alpha_S)$ includes all-order collinear and hard emissions.
They can be expanded as a series in $\alpha_S$ as:
\begin{equation}\label{aas}
A(\alpha_S) \, = \, \sum_{n=1}^{\infty}\left({{\alpha_S}\over
{\pi}}\right)^n A^{(n)}, 
\end{equation}
\begin{equation}
B(\alpha_S) \, = \, \sum_{n=1}^{\infty}\left({{\alpha_S}\over
{\pi}}\right)^n B^{(n)}.
\end{equation}
In the NLL approximation, one needs to include the first two coefficients
of $A(\alpha_S)$ and the first of $B(\alpha_S)$; to NNLL accuracy, 
$A^{(3)}$ and $B^{(2)}$ are also needed.
The coefficients $A^{(1)}$,   $A^{(2)}$ and 
$B^{(2)}$ can be found in \cite{ct}; more recent is the
calculation of the NNLL contributions $A^{(3)}$ \cite{moch1} and
$B^{(2)}$ \cite{moch2}.

Likewise, the threshold-resummed initial condition reads \cite{cc}:
\begin{equation}
\label{deltad}
\Delta^{(D)}_N \left[ \alpha_S(\mu_{0R}^2), \mu_{0R}^2,\mu_{0F}^2, m_c^2
 \right] \, = \, 
\exp
\left\{
G^{(D)}_N \left[ \alpha_S(\mu_{0R}^2), \mu_{0R}^2,\mu_{0F}^2, m_c^2 \right]
\right\},
\eeq
where
\begin{eqnarray}
\label{resdini}
G^{(D)}_N\left[ \alpha_S(\mu_{0R}^2), \mu_{0R}^2, \mu_{0F}^2, 
m_c^2 \right] \, &=& \,
\int_0^1 {dz {{z^{N-1}-1}\over{1-z}}}
\left\{\int^{\mu_{0F}^2}_{m_c^2 (1-z)^2} {{dk^2}\over {k^2}}
A\!\left[\alpha_S(k^2)\right]\right.\nonumber\\  
&+& \left.D\!\left[\alpha_S\left(m_c^2(1-z)^2\right)\right]\right\},
\end{eqnarray}
with $k^2$ and $z$ defined as in (\ref{GNC}).
To NNLL accuracy, we need $A^{(1)}$, $A^{(2)}$ and $A^{(3)}$
and the first two coefficients of 
 \begin{equation}
D(\alpha_S) \, = \, \sum_{n=1}^{\infty}\left({{\alpha_S}\over
{\pi}}\right)^n D^{(n)},
\end{equation}
namely $D^{(1)}$ and $D^{(2)}$.
Function $D(\alpha_S)$, called $H(\alpha_S)$ in \cite{cc},
is characteristic of the fragmentation of 
heavy quarks and resums soft and large-angle
radiation. Its ${\cal O}(\alpha_S)$ 
coefficient can be found in \cite{cc}, 
while $D^{(2)}$ can be read from the formulas in \cite{march,gardi1,alex1}.
In any case, all relevant NNLL threshold-resummation coefficients
are reported in \cite{bfrag}.

In the phenomenological analysis of \cite{bfrag}, the inclusion of
NNLL effects, and especially the contribution  $\sim\alpha_S^3 A^{(3)}$
to function $A(\alpha_S)$, turned out
to be necessary to reproduce the $b$-fragmentation data.
In fact, as we shall point out in the next section, when using
the effective coupling constant we need to redefine the
threshold-resummation coefficients from the third order on.
This way, it turns out that $A^{(3)}$ gets enhanced.
The inclusion
of NNLL terms in the resummed exponents 
shifted the $B$-hadron spectrum towards lower $x_B$ values and played
a crucial role to obtain a 
reasonable description of LEP and SLD data 
(see figure 4 in ref.~\cite{bfrag}).

As in \cite{afr,bfrag}, the Mellin transforms of our
resummed expressions will be performed exactly and
not according to the step-function approximation, which was instead employed 
in the resummations carried out in refs.~\cite{cc,ct}
\footnote{In \cite{cc,ct}, the longitudinal-momentum
integration is done after performing the replacement 
$x^N\to 1-\Theta\left( 1 - z - {{e^{-\gamma_E}}\over N}\right)$,
which is a correct approximation to NLL accuracy.
Beyond NLL, it can be generalized 
following the prescription presented in \cite{deflo}.}.
In fact, as we shall discuss later, we will model non-perturbative
effects to charm fragmentation by means of an effective coupling constant
and it was found in \cite{agl} that
the step-function approximation would 
suppress most power corrections included in the physical observables
via the analytic coupling.
In any case, as thoroughly detailed in \cite{bfrag}, the issue of
the power corrections which are transferred to the 
cross section by the effective coupling, 
and whether it is a better approximation performing
the Mellin transforms in an exact or approximated way is currently
an open issue and we cannot draw any firm conclusion.
A careful analysis, along the lines of \cite{braun}, will be
anyway very welcome to clarify this point.
For the time being, the exactness of the Mellin transforms 
should be seen as part of our non-perturbative model.  
We just point out that, unlike refs.~\cite{cc,ct}, where only
logarithms of $N$ are resummed, in our approach
even some constants and power-suppressed 
${\cal O}(1/N)$ terms are included in the exponents (\ref{rescf}) and
(\ref{deltad}) thanks to the exact Mellin transforms. 
This implies that, any time we improve the accuracy of the
large-$x$ resummation, e.g., from NLL to NNLL, we
include in
the resummed exponent not only subleading logarithms of $N$, but also
constants and power corrections.
This tight relation between perturbative and non-perturbative corrections
is indeed a peculiar feature of our effective-coupling model.

As in \cite{bfrag}, the resummed results are matched to the exact NLO
coefficient function and initial condition. A difference
with respect to the approach followed in 
ref.~\cite{bfrag} is that 
we implement the so-called $\ln R$-matching \cite{logr,gardi}, 
corresponding to matching the logarithms of resummed and NLO expressions. 
We briefly review this matching strategy and how it compares with
the standard method implemented in \cite{bfrag}.
Referring, e.g., to the coefficient function, matched to the
exact NLO one,
it can be written (see eq.~(4.2) in ref.~\cite{bfrag}) as:
\begin{eqnarray}
\label{cfin}
C_N^{'\mathrm{res}}\left[\alpha_S(\mu_R^2), \mu_R^2, \mu_F^2, Q^2 \right]  
&=& K^{(C)}\left[ \alpha_S(\mu_R^2), \mu_R^2, \mu_F^2, Q^2 \right] 
\Delta_N^{(C)}\left[ \alpha_S(\mu_R^2), \mu_R^2, \mu_F^2, Q^2 \right] 
\nonumber\\
&+& d_N^{(C)}\left[ \alpha_S(\mu_R^2), \mu_R^2, \mu_F^2, Q^2 \right].
\end{eqnarray}
In (\ref{cfin}),
$\Delta_N^{(C)}$
is the resummed coefficient function, given in eq.~(\ref{rescf}),
\beq
K^{(C)}\left[ \alpha_S(\mu_R^2), \mu_R^2, \mu_F^2, Q^2 \right] =
1+\alpha_S(\mu_R^2)Q(\mu_F^2,Q^2)\eeq
is
a hard factor, including the constant terms
which are present in the NLO coefficient function but are not resummed in
$\Delta_N^{(C)}$,
\beq
d_N^{(C)}\left[ \alpha_S(\mu_R^2), \mu_R^2, \mu_F^2, Q^2 \right] =
\alpha_S(\mu_R^2)Y(\mu_F^2,Q^2)\eeq
is a remainder function, collecting the left-over NLO terms, suppressed
at large $N$. The explicit expression for functions
$Q(\mu_F^2,Q^2)$ and $Y(\mu_F^2,Q^2)$ can be 
read from the formulas in \cite{bfrag}. A similar expression holds
for the resummed initial condition matched to the NLO result
(see eq.~(5.17) in ref.~\cite{bfrag}).

According to the $\ln R$-matching,
functions $K^{(C)}$ and $d^{(C)}$ are to be replaced by exponential
functions of their ${\cal O}(\alpha_S)$ 
terms and eq.(\ref{cfin}) should read:
\begin{eqnarray}
\label{clogr}
C_N^{\rm res}
\left[\alpha_S(\mu_R^2), \mu_R^2, \mu_F^2, Q^2 \right]  
&=& \exp[\alpha_S(\mu_R^2)Q(\mu_F^2,Q^2)]\times
\Delta_N^{(C)}\left[ \alpha_S(\mu_R^2), \mu_R^2, \mu_F^2, Q^2 \right] 
\nonumber\\ 
&\times& \exp[\alpha_S(\mu_R^2)Y(\mu_F^2,Q^2)].
\end{eqnarray}
From eq.~(\ref{clogr}), one can easily check that the 
logarithms of NLO and resummed functions are actually matched.
In particular, eq.~(\ref{clogr}) differs from (\ref{cfin}) only by terms of
${\cal O}(\alpha_S^2)$ or higher, but it is smoother
at small and large values of $N$ ($x$), thanks to the
exponential functions in eq.~(\ref{clogr}).
It was in fact pointed out in \cite{bfrag} that, since the remainder
function contains terms $\sim\ln x$ and $\sim\ln(1-x)$, the 
physical differential cross sections exhibit oscillating behaviour
near $x\simeq 0$ and $x\simeq 1$. Exponentiating the ${\cal O}(\alpha_S)$ 
contributions to the remainder function should therefore
improve the prediction for small and large values of $x$.
The $\ln R$-matching prescription will be adopted in the following
even for the
resummed initial condition of the perturbative fragmentation function.

The $c$-quark spectrum will finally read in $N$-space as follows:
\begin{eqnarray}
\label{sigmacn}
\sigma^c_N
\!\left[\alpha_S(\mu_{0R}^2),  \alpha_S(\mu_R^2), \mu_{0R}^2, 
\mu_R^2, \mu_{0F}^2, \mu_F^2, m_c^2, Q^2 
\right]\!\!\!
& = &\! C_N^{\rm res} 
\!\left[ \alpha_S(\mu_R^2), \mu_R^2, \mu_F^2, Q^2 \right]\!\nonumber\\ 
&\times&\!
E_N\!\left[ \alpha_S(\mu_{0F}^2),\alpha_S(\mu_F^2) \right]\!
\!\\
&\times&\! 
D_N^{\mathrm{ini,res}}
\left[ \alpha_S(\mu_{0R}^2), \mu_{0R}^2, \mu_{0F}^2, m_c^2 \right]\nonumber.
\end{eqnarray}
In eq.~(\ref{sigmacn}), 
$E_N\!\left[ \alpha_S(\mu_{0F}^2),\alpha_S(\mu_F^2) \right]$ is the
DGLAP operator 
for an evolution between the scales $\mu_{0F}$ and $\mu_F$.
Throughout this paper, we shall implement 
$E_N\!\left[ \alpha_S(\mu_{0F}^2),\alpha_S(\mu_F^2) \right]$
in the non-singlet approximation; 
its explicit expression can be found, e.g., in ref.~\cite{mele}.

\sect{Effective coupling constant}

We shall include non-perturbative corrections to charm fragmentation
using, as in \cite{bfrag}, a  model, based on an extension
of refs.~\cite{shirkov,stefanis}, which 
includes power corrections via an effective strong coupling constant, and
does not introduce any further parameter to be tuned to experimental data.
We review below the main points of our model.

As discussed in ref.~\cite{amati}, in 
resummed calculations the momentum-independent coupling constant is 
replaced by the following integral over the discontinuity of the gluon 
propagator: 
\begin{equation}
\alpha_S\to \frac{i}{2 \pi} \, \int_0^{k^2} d s 
\ {\rm Disc}_s\  \frac{\alpha_S(-s) }{ s },
\label{ask}
\end{equation}
where $k^2$ is the gluon transverse momentum relative to the emitter,
defined, e.g., as in eq.~(\ref{GNC}). 
In eq.~(\ref{ask}) the discontinuity is given by:
\begin{equation}
{\rm Disc}_s F(s) =  \lim_{\epsilon \, \to \, 0^+} 
\left[F(s+i\epsilon)- F(s-i\epsilon)\right].
\end{equation}
At LO, e.g., $\alpha_S(-s)$ reads:
\begin{equation}\label{ipi}
\alpha_{S,\mathrm{LO}}(-s)=\frac{1}{\beta_0[\ln(|s|/\Lambda^2)-i\pi
\Theta(s)]},
\end{equation}
where $\beta_0= (33-2n_f)/(12\pi)$ is the first-order term of the
QCD $\beta$-function, $n_f$ is number of active flavours, and
$\Lambda$ is the QCD scale, e.g., in the \msbar renormalization scheme.

The integral (\ref{ask}) is usually carried out neglecting the imaginary part,
$\sim i\pi$, in the denominator of $\alpha_S(-s)$, 
i.e. assuming
\begin{equation}\label{ppi}
\ln{{|s|}\over{\Lambda^2}}\gg \pi
\end{equation}
in eq.~(\ref{ipi}).
The approximation (\ref{ppi}) allows one to avoid the Landau pole, so
that the integral (\ref{ask}) turns out to be
roughly equal to 
the strong coupling constant evaluated at the upper integration limit:
\begin{equation}
\frac{i}{2 \pi} \, \int_0^{k^2} d s 
\ {\rm Disc}_s\  \frac{\alpha_S(-s) }{ s }\, \simeq\, \alpha_S(k^2).
\label{assk}
\end{equation}
In fact, resummed formulas typically use the transverse momentum $k^2$ as 
the scale of the strong coupling constant \cite{ct}.

As in \cite{bfrag}, we shall follow a different approach and 
avoid the Landau pole by using in eq.~(\ref{ask})
a regularized coupling constant 
$\bar \alpha_S$, defined as follows \cite{shirkov,afr}:
\begin{equation}
\bar\alpha_S(k^2)= 
\frac{1}{2\pi i}
\int_0^{\infty}  \, \frac{ds}{s+k^2} \, 
{\rm Disc}_s \, \alpha_S(-s).
\label{space}
\end{equation}
Inserting in (\ref{space}) the LO expression (\ref{ipi})
and performing the integration, we obtain:
\begin{equation}
\bar\alpha_{S,\mathrm{LO}}
(k^2) \, = \,{1\over{\beta_0}}\left[{1\over{\ln(k^2/\Lambda^2)}}-
{{\Lambda^2}\over{k^2-\Lambda^2}}\right].
\label{spacelo}
\end{equation}
If we compare eq.~(\ref{spacelo}) with the LO standard coupling, i.e.
\begin{equation}
\alpha_{S,\mathrm{LO}}(k^2) \, = \, {1\over{\beta_0\ln(k^2/\Lambda^2)}},
\label{as}
\end{equation}
we learn that in eq.~(\ref{spacelo}) a power-suppressed term, 
relevant at small $k^2$, has subtracted
off the Landau pole $k^2=\Lambda^2$, 
which is instead present in (\ref{as}). 
At large $k^2$, $\bar\alpha_S(k^2)$ is nonetheless still roughly equal to
$\alpha_S(k^2)$. Such results can be generalized to higher 
accuracy levels, using the two- and three-loop beta function, 
as done in \cite{bfrag}.

The effective coupling constant $\tilde\alpha_S(k^2)$ will be still
defined as in eq.(\ref{ask}), but using the analytic coupling (\ref{space})
in the integrand function: 
\begin{equation}
\tilde\alpha_S(k^2) = \frac{i}{2 \pi} \, \int_0^{k^2} d s 
\ {\rm Disc}_s\  \frac{ \bar\alpha_S(-s) }{ s }.
\label{time}
\end{equation}
Using the LO result (\ref{spacelo}), we can perform the
integral (\ref{time}) and obtain our LO effective coupling constant:
\beq
\tilde\alpha_{S,\mathrm{LO}}(k^2)={1\over\beta_0}\left\{{1\over 2}-{1\over\pi}
\arctan\left[{{\ln(k^2/\Lambda^2)}\over{\pi}}\right]\right\}.
\label{atan}
\eeq
The NLO and NNLO expressions of $\tilde\alpha_S(k^2)$ can be found
in \cite{bfrag}. 
It is straightforward to show that eq.~(\ref{atan}), as well as its
higher-order generalizations, 
is free from the Landau pole and includes 
power-suppressed contributions at small momenta.
Also, as discussed in \cite{agl}, eq.~(\ref{time}) accounts for 
absorptive effects due to gluon branching, since we are not neglecting
any longer the imaginary part in the denominator of $\alpha_S(-s)$.

In principle, both analytic coupling constants (\ref{space}) and
(\ref{time}) are possible 
candidates to model
non-perturbative corrections \footnote{In the literature \cite{bfrag,shirkov},
one usually
refers to $\bar\alpha_S(k^2)$ and $\tilde\alpha_S(k^2)$ as effective
space- and time-like coupling constants, respectively.}.
However, as debated in \cite{bfrag}, it is only
(\ref{time}) which gives an acceptable 
description of $b$-fragmentation data and we shall therefore stick to 
$\tilde\alpha_S(k^2)$ to model power corrections to charm fragmentation
as well.

The relation between effective and standard coupling constant
for $\ln(k^2/\Lambda^2)\gg \pi$ reads:
\begin{equation}
\tilde\alpha_S(k^2) \, = \, \alpha_S(k^2) 
\, - \, \frac{\left(\pi\beta_0\right)^2}{3} \, \alpha_S^3(k^2)
\, + \, {\cal O}(\alpha_S^4).
\label{atas}
\end{equation}
From eq.~(\ref{atas}) we learn that
at high energy 
the difference between $\tilde\alpha_S(k^2)$
and $\alpha_S(k^2)$ starts from ${\cal O}(\alpha_S^3)$.
Moreover, eq.~(\ref{atas}) dictates that, when employing the
effective coupling constant, 
we will have to redefine the soft-resummation coefficients from 
order $\alpha_S^3$ on. As anticipated in subsection~\ref{sub2}, 
the NNLL coefficient $A^{(3)}$ of the ${\cal O}(\alpha_S^3)$ term of function
$A(\alpha_S)$, entering in Eqs.~(\ref{GNC}) and
(\ref{resdini}), will get enhanced according to:
\begin{equation}
A^{(3)} \, \to \, \tilde A^{(3)} \, = \, A^{(3)} \, + 
\, {{(\pi\beta_0)^2}\over 3}A^{(1)}.
\label{a3}
\end{equation}

The other assumptions contained in our model are 
also detailed in ref.~\cite{bfrag}
and we do not report them here for the sake of brevity.
We just point out that, when dealing with higher orders of 
$\tilde\alpha_S(k^2)$,
 we shall adopt the so-called `power-expansion' choice, which implies that
we shall evaluate the powers $\tilde\alpha_S^n(k^2)$
after computing the integral over the discontinuity:
\beq
\label{power_exp}
\tilde{\alpha}_S^n(k^2)=
\left[\frac{i}{2 \pi} \, \int_0^{k^2} d s 
\ {\rm Disc}_s\  \frac{ \bar\alpha_S(-s) }{ s }\right]^n.
\eeq
On the contrary, the original proposal in \cite{shirkov}
consisted in calculating the discontinuity of $\bar\alpha_S^n(-s)$
before integrating over $s$ (`non power-expansion' choice). As discussed in
\cite{bfrag}, the non-power expansion prescription would yield a rather
poor description of $b$-fragmentation data.

The purpose of the present paper is indeed to push 
the effective-coupling model to lower energies and compare its
predictions with data on $c$-flavoured hadron production.
For a consistent comparison with the results obtained in the
framework of $B$-hadron production and decay, throughout this paper, as in 
\cite{afr,bfrag}, we shall use
$\tilde\alpha_S(k^2)$ evaluated to three-loop accuracy, 
everywhere in our calculation, i.e. in
both coefficient function
and perturbative fragmentation function. Hereafter, the effective
coupling constant (\ref{time}) will be our only source of non-perturbative
corrections and we shall not introduce any further non-perturbative
fragmentation function.

It was pointed out in \cite{bfrag}
that power-correction effects in the initial condition of the perturbative
fragmentation function are more relevant than in the coefficient function.
The typical $c$-fragmentation
scales at which the coupling constant is evaluated
are, in fact, $C=Q\sqrt{1-x}$ in the coefficient function and 
$S=m_c(1-x)$ in the initial condition. $C$ and $S$ are the
integration limits in the resummed exponents as
well as the arguments of $\alpha_S$
in functions $B\left[\alpha_S(C^2)\right]$ and $D\left[\alpha_S(S^2)\right]$, 
appearing in the large-$x$ resummation expressions 
(\ref{GNC}) and (\ref{resdini}).
If we calculate $C$ and $S$ for $Q=m_Z$ and
 $x=0.5$, where, as will be shown in the next section, the $D$-meson 
spectrum in $e^+e^-$ annihilation is roughly
peaked, we shall get $C\simeq 46$~GeV, $S\simeq 0.9$~GeV,
$\tilde\alpha_S(C^2)\simeq 0.13$ and $\tilde\alpha_S(S^2)\simeq
0.35$. Therefore, non-perturbative corrections are more important in
the initial condition, depending on $S$, than in the coefficient function.
Comparing now 
the values of $\tilde\alpha_S$ at the charm- and bottom-mass scales, 
we find that $\tilde\alpha_S (m_c^2)\simeq 0.3$ is 
appreciably higher than  
$\tilde\alpha_S (m_b^2)\simeq 0.2$.
However, it is interesting to
notice that the scales $S$ and $C$, and hence $\tilde\alpha_S(S^2)$ and
$\tilde\alpha_S(C^2)$, are roughly the same for bottom and charm production
if evaluated at the maxima of the respective spectra at LEP, i.e.
$x$=0.5 for $D$- and $x=0.8$ for $B$-hadron energy distributions.

Before closing this section, we would like to
stress that, in its current formulation,
our parameter-free model works in 
the same fashion for $B$ as well $D$ mesons, up to the replacement
$m_b\to m_c$. Also, our model does not distinguish among 
baryons and mesons, spin-1 and spin-0, charged and neutral hadrons.
It was therefore argued in \cite{bfrag} that possible extensions of our
model may consist in including a correcting term, so that 
\beq
\label{dal}
\tilde\alpha_S(k^2)\to \tilde\alpha_S(k^2)+\delta\tilde\alpha_S(k^2)\ ,\eeq
where
$\tilde\alpha_S(k^2)$ is still the effective coupling discussed
above, and $\delta\tilde\alpha_S(k^2)$ may depend, e.g., on whether
we have baryons or mesons, $B$'s or $D$'s, and so on.
The analysis which we shall undertake 
herafter should therefore be helpful
to establish, for the time being, whether the contribution  
$\delta\tilde\alpha_S(k^2)$ 
is mandatory or not.

\sect{Results in $x$-space}

In this section we compare our results in $x$-space with experimental
data on $c$-flavoured hadron production in $e^+e^-$ annihilation.
Hadronization effects will be accounted for by employing the
analytic coupling constant (\ref{time}) at NNLO.
Whenever we use 
$\tilde\alpha_S(k^2)$ instead of the standard $\alpha_S(k^2)$, 
the charm-quark energy fraction will be replaced by its hadron-level
counterpart:
\beq x_D={{2p_D\cdot Q}\over{Q^2}},\eeq with $p_D$ being the momentum
of a $D$-hadron.
The $D$  spectrum in moment space will be written
in a form analogous to eq.(\ref{sigmacn}), up to $\alpha_S\to \tilde\alpha_S$:
\begin{eqnarray}
\label{sigmabhad}
\sigma^{(D)}_N( \mu_R^2, \mu_{0R}^2, \mu_{0F}^2, \mu_F^2, m_c^2, Q^2 )
&=& C_N^{\rm res}
\left[ \tilde\alpha_S(\mu_R^2), \mu_R^2, \mu_F^2, Q^2 \right] \times\
E_N\left[ \tilde{\alpha}_S(\mu_{0F}^2), \tilde{\alpha}_S(\mu_F^2) \right]
\nonumber\\
&\times &
D_N^{\mathrm{ini,res}} 
\left[ \tilde{\alpha}_S(\mu_{0R}^2), \mu_{0R}^2, \mu_{0F}^2,m_c^2 \right].
\end{eqnarray}
The $x$-space result is then recovered by performing an inverse Mellin
transform:
\beq\label{inverse}
\sigma^{(D)}
\left( x_D; \,  \mu_R^2, \mu_{0R}^2, \mu_{0F}^2, \mu_F^2, m_c^2, Q^2 \right)
\, = \, \int_{\gamma-i\infty}^{\gamma+i\infty}
\frac{dN}{2\pi i} x_D^{-N} 
\, \sigma^{(D)}_N( \mu_R^2, \mu_{0R}^2, \mu_{0F}^2, \mu_F^2, m_c^2, Q^2 ),
\eeq
where $\gamma$ is a positive constant.
As discussed in \cite{bfrag}, since the effective $\tilde\alpha_S(k^2)$
does not exhibit the Landau pole any longer, we do not need any prescription,
such as the well-known
minimal prescription \cite{min}, to avoid the Landau pole
in the integration  (\ref{inverse}). The integral will be performed
in a numerical way, along the lines of \cite{bfrag}; it was checked
that the results are stable when varying the integration contour, i.e.
the constant $\gamma$.

As in ref.~\cite{cno}, 
we shall consider LEP data from the ALEPH collaboration \cite{lep}, taken
at the $Z^0$ pole, and data from the
CLEO \cite{cleo} and BELLE \cite{belle}
experiments, at the $\Upsilon(4S)$ resonance.
We shall investigate neutral as well as charged
$D$ and $D^*$ mesons; 
in fact, we just pointed out that our model does not
distinguish the hadron electric charge or spin.

As discussed in \cite{cno}, electromagnetic initial-state radiation (ISR)
effects can modify the shape of charmed-meson spectra. Such effects
are important especially at $B$-factories, where 
the emission of photons from the $e^+e^-$ pair, whose rate is
$\sim\alpha\ln(Q^2/m_e^2)$, $m_e$ being the electron mass, 
may significantly decrease the
energy in the centre-of-mass system. The CLEO and BELLE data did not account
for such effects, which were instead 
implemented in the analysis \cite{cno}. 
In the following, we shall compare with data corrected for ISR effects: 
a discussion on the impact of such contributions on
$D$-spectra in $x$- and $N$-spaces can be found in \cite{cno}.
Such effects were also implemented to correct the ALEPH data,
but it was understood that at the $Z^0$ pole they are quite negligible.


The non-perturbative model based on the effective coupling constant
(\ref{time}) does not have any free parameter to be tuned to the data
which we shall consider.
We shall nonetheless vary the parameters entering in
the perturbative calculation in such a way to give an estimate of the
theoretical uncertainty on our prediction.
We change each quantity separately, keeping the others to their default
values, in such a way to avoid too many runs.

Following \cite{bfrag}, the default values of our perturbative parameters
will be
$\mu_R=\mu_F=Q$ and $\mu_{0R}=\mu_{0F}=m_c$,
where $\mu_R$ and $\mu_F$ are the renormalization
and factorization scales in the coefficient function, and 
$\mu_{0R}$ and $\mu_{0F}$ in the initial condition of the perturbative
fragmentation function.
The hard scale will be $Q=m_Z$ or $m_{\Upsilon(4S)}$ at LEP or $B$-factories,
with $m_Z=91.19$~GeV and $m_{\Upsilon(4S)}=10.58$~GeV.
We shall vary $\mu_R$ and $\mu_F$
between $Q/2$ and $2Q$, $\mu_{0R}$ and $\mu_{0F}$ between
$m_c/2$ and $2m_c$.
As in \cite{bfrag}, we shall let
$\alpha_S(m_Z^2)$ run
in the range $0.117<\alpha_S(m_Z^2)<0.121$, using
$\alpha_S(m_Z^2)=0.119$ as our default value.
The corresponding variation range of the effective coupling constant is
$0.115<\tilde\alpha_S(m_Z^2)<0.119$.
For the purpose of $m_c$, as thoroughly discussed in \cite{bfrag}, using the
pole or the \msbar heavy-quark mass definition in the initial condition
is equivalent for calculations relying on the NLO/NLL approximation.
However, 
in the NNLL large-$x$ resummation of the initial condition,
and in particular in the definition of the coefficient
$D^{(2)}$ in eq.~(\ref{resdini}),
we are employing results of the  NNLO
computation in \cite{alex1}, which uses the heavy-quark pole mass.
Hence, we should use the charm pole mass as well.
Nonetheless, as pointed out
in \cite{bfrag}, when we use the effective coupling constant to
describe hadronization corrections, it is not uniquely determined whether
$m_c$ should be the quark or the hadron mass. 
As done for the purpose
of the bottom-quark mass, we shall adopt a
conservative choice and vary $m_c$ in the range 
$1.5~\mathrm{GeV} \, < \, m_c \, < \, 2.1~\mathrm{GeV}$, that includes the
current estimations for the charm pole mass as well as $D$-hadron masses
\cite{pdg}. Our default value will be $m_c$=1.8 GeV.

As for the DGLAP evolution operator, when evolving from 
$\mu_{0F}\simeq m_c$ to $\mu_F \simeq Q$, one typically crosses the 
bottom-quark mass threshold $m_b$. ref.~\cite{cno1} computed
at NLO the matching conditions  for the perturbative fragmentation
function when crossing heavy-flavour thresholds.
In our study, however, since we are working in the
non-singlet approximation and we are not accounting for gluon splitting
and flavour mixing, we shall neglect such matching conditions.
In fact, we checked that our results change very little
according to whether we set in the DGLAP evolution
operator, e.g, $n_f=4$ or $n_f=5$ as the number of active flavours.
In any case, in our phenomenological analysis, whenever we have
$\mu_{0F}<m_b<\mu_F$, 
we shall implement
the following factorized form for the non-singlet DGLAP evolution operator:
\begin{equation}\label{dglap}
E_N\left[\tilde\alpha_S(\mu_{0F}^2),
\tilde\alpha_S(\mu_F^2)\right]=E_N\left[\tilde\alpha_S(\mu_{0F}^2),
\tilde\alpha_S(m_b^2)\right]\times E_N\left[\tilde\alpha_S(m_b^2),
\tilde\alpha_S(\mu_F^2)\right],
\end{equation}
with $n_f=4$ and $n_f=5$ below and above the bottom-quark mass threshold,
respectively.
The $b$-quark mass will be varied in the range 
$4.7~\mathrm{GeV} \, < \, m_b \, < \, 5.3~\mathrm{GeV}$, as in \cite{bfrag},
with $m_b=5$~GeV being our default value.
Elsewhere in our calculation, $n_f$ will be consistently
chosen according to the energy scale we are dealing with.

\subsection{Comparison with ALEPH data}

We shall first consider ALEPH data on ${D^*}^+$ production. As detailed
in \cite{lep}, such mesons can be in general produced from a $Z^0\to c\bar c$
decay, from the decay of a primary
$b$-flavoured hadron produced in $Z^0\to b\bar b$,
from gluon splitting to $c\bar c$ or $b\bar b$ pairs, which subsequently
hadronize or decay into a ${D^*}^+$. The ALEPH Collaboration was 
able to subtract the $Z^0\to b\bar b$ and gluon-splitting
contributions off 
and published the spectrum of ${D^*}^+$ mesons coming only from
the $c\bar c$ primary source. 
In the following, we shall compare the predictions of
our model with such a subsample, which will allow us to
neglect secondary charm production in the perturbative calculation
as well as the singlet component of the DGLAP evolution operator, which in 
principle should play a role at LEP energies.
Indeed, it was found out in \cite{cno}
that implementing the singlet contribution does have an 
effect at $x_D<0.4$, and 
actually worsens the comparison with the ALEPH ${D^*}^+$ data
coming from direct $c\bar c$ production (see 
figures 13 and 14 in ref.~\cite{cno}).

In figure \ref{aleph} we present the spectrum given by our model, along
with the ${D^*}^+$
ALEPH data, and investigate the dependence
on the factorization scales $\mu_F$ and $\mu_{0F}$ (figure \ref{aleph} (a)), 
and on the choice of $\alpha_S(m_Z^2)$ and  $m_c$
(figure \ref{aleph} (b)). For the sake of comparison,
both data and theoretical predictions are normalized to unity.
As already observed in \cite{bfrag}, the dependence on $\mu_{0F}$,
the scale entering in the initial condition of the perturbative fragmentation
function, is fairly large, while the impact of the choice of $\mu_F$ is pretty 
small. In particular, setting a lower value of $\mu_{0F}$, 
e.g. $\mu_{0F}=m_c/2$, tends to
deplete the small-$x_D$ region of the spectrum and to
enhance the event fraction 
around the peak. Also, the peak is slightly shifted to higher
$x_D$ if we choose $\mu_{0F}=m_c/2$.
The prediction obtained for $\mu_{0F}=2m_c$ reproduces quite well 
the low-$x_D$ data, while discrepancies are still present in the
middle-high range.

The dependence on $\alpha_S(m_Z^2)$ and $m_c$ is also quite
relevant, as can be learned from figure \ref{aleph}~(b).
In particular, a low value of $m_c$, 
i.e. $m_c=1.5$~GeV, consistent with the quark mass rather than 
the ${D^*}^+$-meson mass, gives a pretty good description of the peak,
but it worsens the comparison for $x_D>0.7$.
On the contrary, a high value of $m_c$, such as 2.1 GeV, significantly
moves the peak towards large $x_D$ and worsens the overall comparison.
As for the effect of the variation of $\alpha_S(m_Z^2)$, we find that
it shifts the position of the peak:
the lower $\alpha_S(m_Z^2)$, the higher the value of
$x_D$ at which the $D$ spectrum is peaked.
The dependence on the renormalization scales $\mu_R$ and $\mu_{0R}$ is 
very little, and we do not present the corresponding
plots for the sake of brevity. 
We also varied $m_b$, the bottom-quark mass entering in eq.~(\ref{dglap}),
but found out that it has negligible
impact on the energy distribution.
\FIGURE{
    \epsfig{file=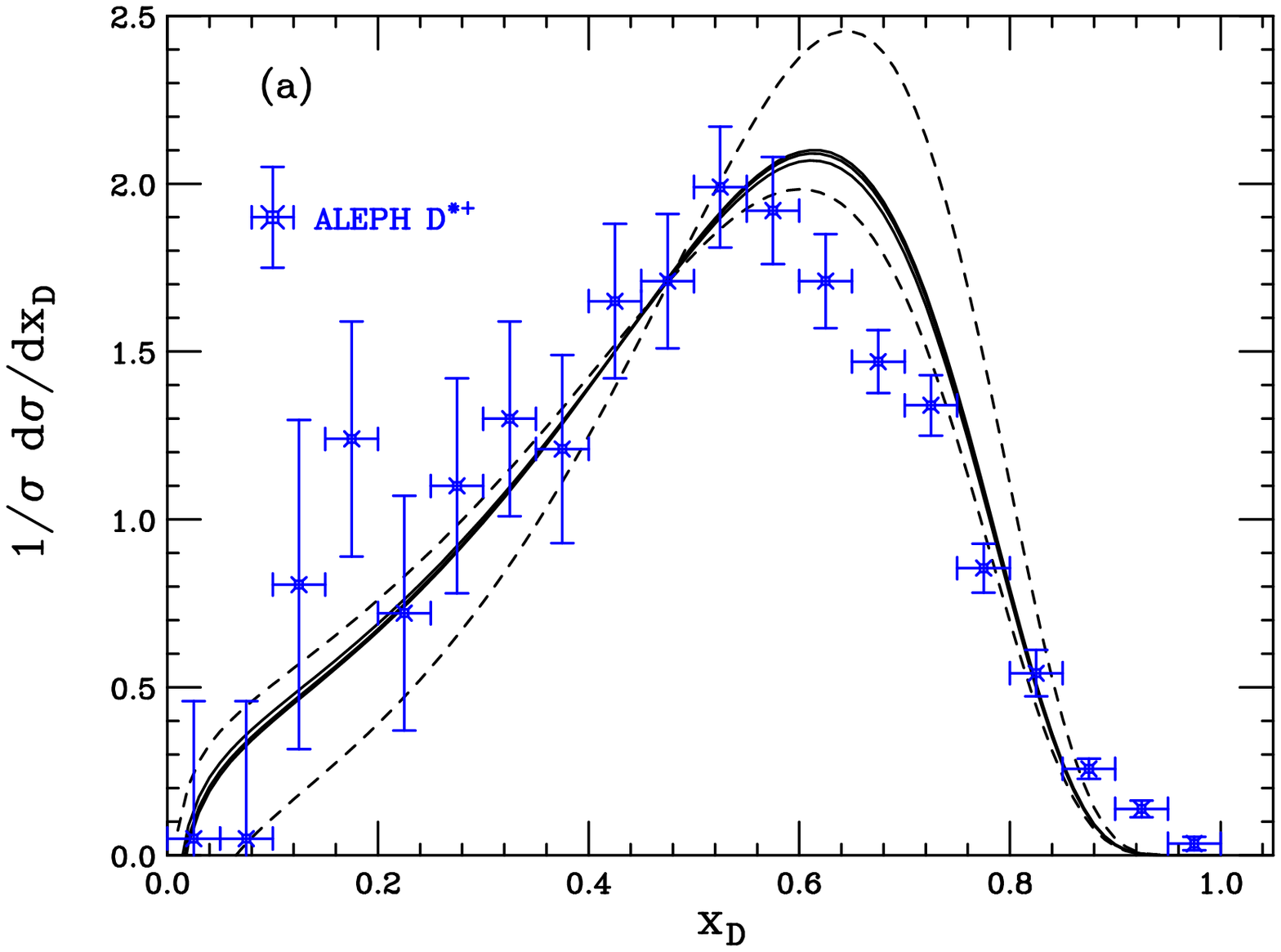,width=0.49\textwidth}
    \epsfig{file=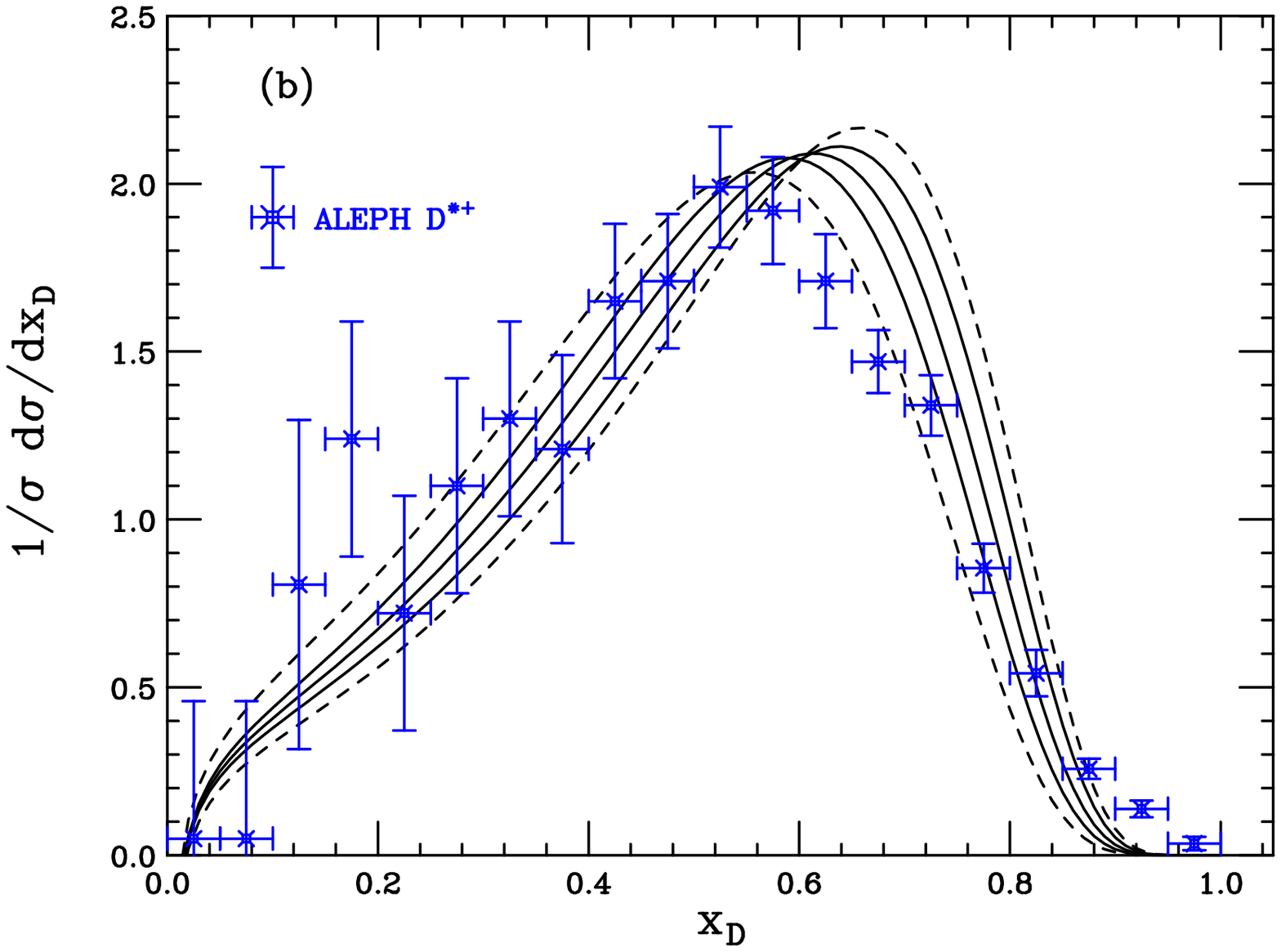,width=0.49\textwidth}
\caption{Comparison of the prediction yielded by our model with data from
ALEPH on ${D^*}^+$ production a LEP. 
In (a) we investigate the dependence 
on $\mu_F$ (solid lines) and $\mu_{0F}$ (dashes); in 
(b) the effect of the choice of $\alpha_S(m_Z^2)$ (solid) and $m_c$ (dashes). 
Such quantities are varied as discussed in the text.}
\label{aleph}}

Overall, we can say that our model
gives an acceptable description of the raise at low and average values of
$x_D$, while discrepancies are present around the peak, unless one
sets a relatively low value for $m_c$, and at very large
$x_D$. Our curves tend to be harder than the data 
and approach zero at large $x_D$ more rapidly.
Although the comparison at very large $x_D$ is not completely satisfactory,
using the $\ln R$-matching prescription,
discussed in subsection~\ref{sub2}., 
has nonetheless improved the spectrum near the endpoint $x_D=1$, 
as it is smoother
and not oscillating any longer.
We checked that if we had used the standard matching between NLO and resummed
expressions as in \cite{bfrag}, the charmed-meson distributions would have
become negative for $x_D\gsim 0.9$.

In any case, we are aware that our model, based on an
extrapolation of perturbation theory, up to the replacement of the
coupling constant $\alpha_S(k^2)\to\tilde\alpha_S(k^2)$, cannot be completely
reliable at very 
large $x_D$.
One can roughly estimate \cite{cc}
$x_{D,\mathrm{max}}
\simeq 1-\Lambda/m_c\simeq 0.85$
the maximum value of $x_D$ at which our model, or any model
based on simple parametrizations of power corrections, such 
as the non-perturbative fragmentation functions \cite{kart,pet},
can be trusted.
In fact, the authors of 
ref.~\cite{cno} managed to improve the comparison at large 
$x_D$, but they had to introduce a further free
 parameter which they tuned to data.
In detail, they replaced the Mellin variable $N$ according to:
\begin{equation}\label{nprime}
N\to N\frac{1+f/N'}{1+fN/N'},\end{equation}
where $N'=\exp[1/\left(\beta_0\alpha_S(\mu_R^2)\right)]$ 
in the coefficient function
and  $N'=\exp[1/\left(2\beta_0\alpha_S(\mu_{0R}^2)\right)]$ 
in the initial condition. ref.~\cite{cno} used then $f=1.25$
in its phenomenological analysis, as
this choice led to good fits to charm-fragmentation data.
In principle, we could also perform the replacement (\ref{nprime})
and tune $f$.
In fact, modifying the energy distribution at large $x_D$
will also have an impact at smaller values of $x_D$, since 
we have kept the normalization of data and theory curves fixed 
to unity. However,
in this way our model
would lose its crucial feature to be free from tunable parameters. 
Furthermore, given the theoretical
uncertainty on our prediction, the value of $f$ will depend on
the particular set of perturbative parameters chosen. Therefore, for the time 
being, we prefer to stick to our parameter-free modelling of the hadronization
and to postpone a more careful investigation of the very large-$x_D$ regime
of our spectra to future work, with the NNLO corrections implemented.
In any case, we should never forget that, for the sake of consistency, 
whenever we modify the perturbative accuracy or the non-perturbative model,
we should always
reconsider the studies on 
$B$-hadron production and decay and check whether the 
results obtained in refs.~\cite{afr}
and \cite{bfrag} still hold.

In the present analysis, as in ref.~\cite{bfrag}, we discard 
few points at very large $x_D$ and limit ourselves
to $x_D\leq 0.85$ when evaluating the $\chi^2$ from the comparison
with the data.
Even in this range, using our default values for the parameters
in the parton-level computation, we are not able to acceptably reproduce
the data, as we obtain $\chi^2/\mathrm{dof}=56.47/17$. 
A better description of the data is nonetheless 
obtained if, e.g., we keep all 
quantities to their default values, but set
$\mu_{0F}=2m_c$ ($\chi^2/\mathrm{dof}=27.18/17$) or 
$\alpha_S(m_Z^2)=0.121$ ($\chi^2/\mathrm{dof}=30.52/17$).
Setting $m_c=1.5$~GeV, we find $\chi^2/\mathrm{dof}=32.29/17$.
As we are not fitting any non-perturbative parameter to the data,
such values of $\chi^2$ are acceptable.
Also, they are of similar magnitude to those
obtained in \cite{bfrag} from the comparison with
$B$-hadron energy distributions at the $Z^0$ pole for $x_B<1-\Lambda/m_b$.

The overall impact of the inclusion of non-perturbative corrections 
at LEP energies via
our model can be learned from figure \ref{pnp2}, where we present our most
significant predictions, 
i.e. the ones obtained with
$\mu_{0F}=2m_c$ and $m_c=1.8$~GeV (solid line), and 
with  $\mu_{0F}=m_c$ and $m_c=1.5$~GeV (dotted), keeping
the other quantities to their default
values. In figure \ref{pnp2} we also show the ALEPH ${D^*}^+$ data
and the purely perturbative results of ref.~\cite{cc}, 
where the authors used the standard coupling constant and resummed
NLL soft and collinear contributions to both coefficient function and
perturbative fragmentation function.
The role played by power corrections is clearly 
remarkable throughout all $x_D$-spectrum, 
and is essential to obtain an acceptable description of the data.
In fact, 
the parton-level calculation of \cite{cc}, which is the same as 
the one employed in \cite{cno}, 
needs to be convoluted with a
non-perturbative fragmentation function to reproduce the data.
We can also note in figure \ref{pnp2} that, while setting $\mu_{0F}=2m_c$
and $m_c=1.8$~GeV
gives the lowest $\chi^2$, the data around the peak are better described if
we instead choose $m_c=1.5$~GeV and $\mu_{0F}=m_c$.

Before closing this subsection, we remind
that the possible reasons 
determining the fairly large theoretical uncertainties
were already listed and detailed 
in \cite{bfrag}. In particular,
we have resummed large-$x$ 
contributions to the coefficient function and initial condition in the
NNLL approximation, but we have still matched the resummation to the
NLO exact results, thus generating a mismatch between the
NNLL terms in the resummed exponents ($\sim\alpha_S^n\ln N$, etc.)
and the remainder functions. We believe that the uncertainties should
be milder if we used the exact NNLO results \cite{neerven,moch,alex1,alex3}.
Moreover, lower theoretical errors should be expected if we also
employed NNLL non-singlet DGLAP evolution equations, using 
NNLO non-singlet splitting functions \cite{alex2}.

With respect to the analysis on $B$-hadron production, the effect of 
the choice of scales and masses is here even more relevant: 
the dependence on such quantities is typically logarithmic, hence
larger once they vary around $m_c$
rather than $m_b$. It is however interesting to notice that, unlike
the comparison with the $B$-hadron data, where setting
$\mu_{0F}=m_b/2$ gave the best description of the data 
\cite{bfrag}, the charm-fragmentation data seem to prefer a quite
high value of $\mu_{0F}$, since $\mu_{0F}=2m_c$ yields the lowest
$\chi^2$. 
We believe that a full NNLO/NNLL analysis should
clarify this issue as well.

\FIGURE{
 \label{pnp2}
\epsfig{file=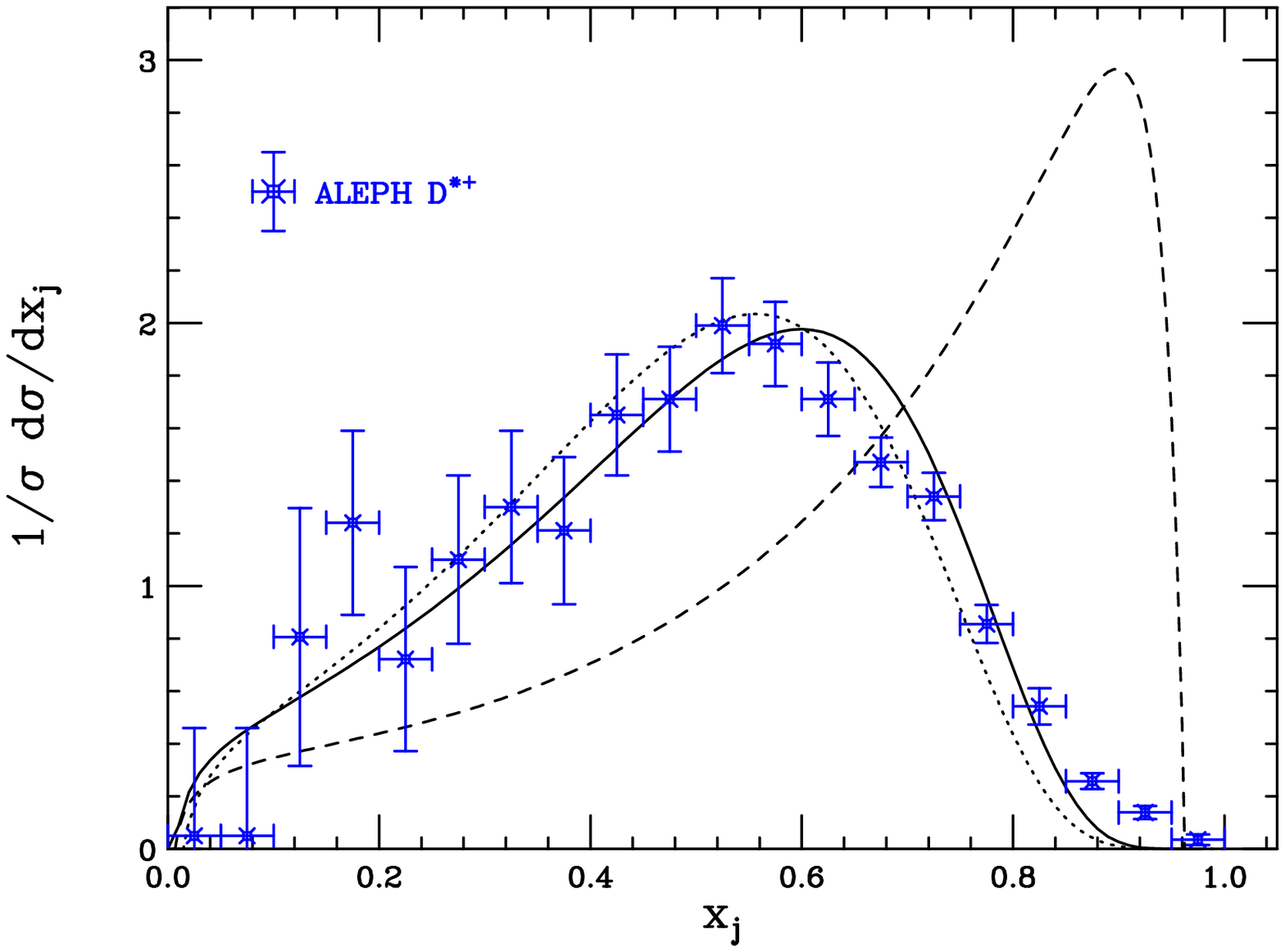, width=.7\textwidth}
\caption{Results on
charmed-hadron production at LEP ($j=D$), setting $\mu_{0F}=2m_c$
and $m_c=1.8$~GeV
(solid), $m_c=1.5$~GeV and $\mu_{0F}=m_c$ (dots),
compared with the perturbative parton-level calculation of \cite{cc}
($j=c$, dashes) and the ALEPH ${D^*}^+$ data.}}


\subsection{Comparison with CLEO and BELLE data}

We would like to compare the predictions of our model with the
data on $D^0$, ${D^*}^0$ and ${D^*}^+$ production 
from the experiments CLEO \cite{cleo} and BELLE \cite{belle},
collected at the $\Upsilon(4S)$ resonance.
In fact, since the value of the
hard scale is much smaller than at LEP, such a comparison
will help
to shed light on the performance of our model and calculation at lower
energies. 
Furthermore, ref.~\cite{cno}, using a NLO/NLL calculation and
a non-perturbative fragmentation function with three parameters, managed
to fit all $B$-factory data, whereas some discrepancies were found with
respect to the ALEPH data after evolving to LEP energies.
Our case is clearly different, as our non-perturbative model is not tunable to
data, but it will be nonetheless cumbersome to investigate how our predictions
fare with respect to the different data sets at the $\Upsilon(4S)$  
resonance and estimate the theoretical uncertainty.



In figure \ref{db} we present the comparison with CLEO and BELLE
data on $D^0$ production, corrected for ISR effects. As pointed out in
\cite{cno}, at the $\Upsilon(4S)$ resonance the contribution of 
$c\bar c$ pair production via gluon splitting is 
negligible, hence it is safe sticking to the non-singlet approximation
of the DGLAP evolution equations, as done for analysis at LEP energies.
The data sets which we consider are separately normalized to 1, for the sake
of a consistent comparison with the theory curves, whose first moment
reads, by definition, $\sigma_{N=1}=1$. We vary 
renormalization and factorization scales, 
$m_c$ and $\alpha_S(m_Z^2)$ along the lines of our comparison with ALEPH.
figure \ref{db} (a) exhibits the dependence on $\mu_F$ and $\mu_{0F}$;
figure \ref{db} (b) the one on $\alpha_S(m_Z^2)$ and $m_c$. We do
not present the effect of changing $\mu_R$, $\mu_{0R}$
and $m_b$, since it is very
little, as already found at the $Z^0$ pole. 

Unlike the comparison with the ALEPH data, where, though within
the experimental and theoretical  
uncertainties, we succeeded in getting a reasonable fit of the data,
our prediction lies quite far from the CLEO and BELLE $D^0$ data 
and there is no
choice of parameters and scales, within our  ranges, which
can accommodate the experimental data. In fact, such data
exhibit very small errors and, 
even if we limit our analysis to $x_D<0.85$, as we did before, we still
obtain quite large $\chi^2$, typically 
$\chi^2/\mathrm{dof}\gsim {\cal O}(10)$. 
It is nonetheless interesting to notice that the best comparison is obtained
for $m_c=1.5$~GeV: in this case, one is able at least to reproduce the rise
of the spectrum
up to $x_D\simeq 0.6$, but still uncapable of describing the peak 
and the large-$x_D$ tail.
As pointed out when comparing with ALEPH, a full NNLO/NNLL analysis is
mandatory to reduce the theoretical error and 
should shed light on the
dependence on the quark (meson) mass as well.
Ref.~\cite{cno} also presented $D^+$ data from
CLEO and BELLE; the comparison with our predictions is however qualitatively
similar to the one presented in figure \ref{db} and we do not show it
for brevity.

\FIGURE{
    \epsfig{file=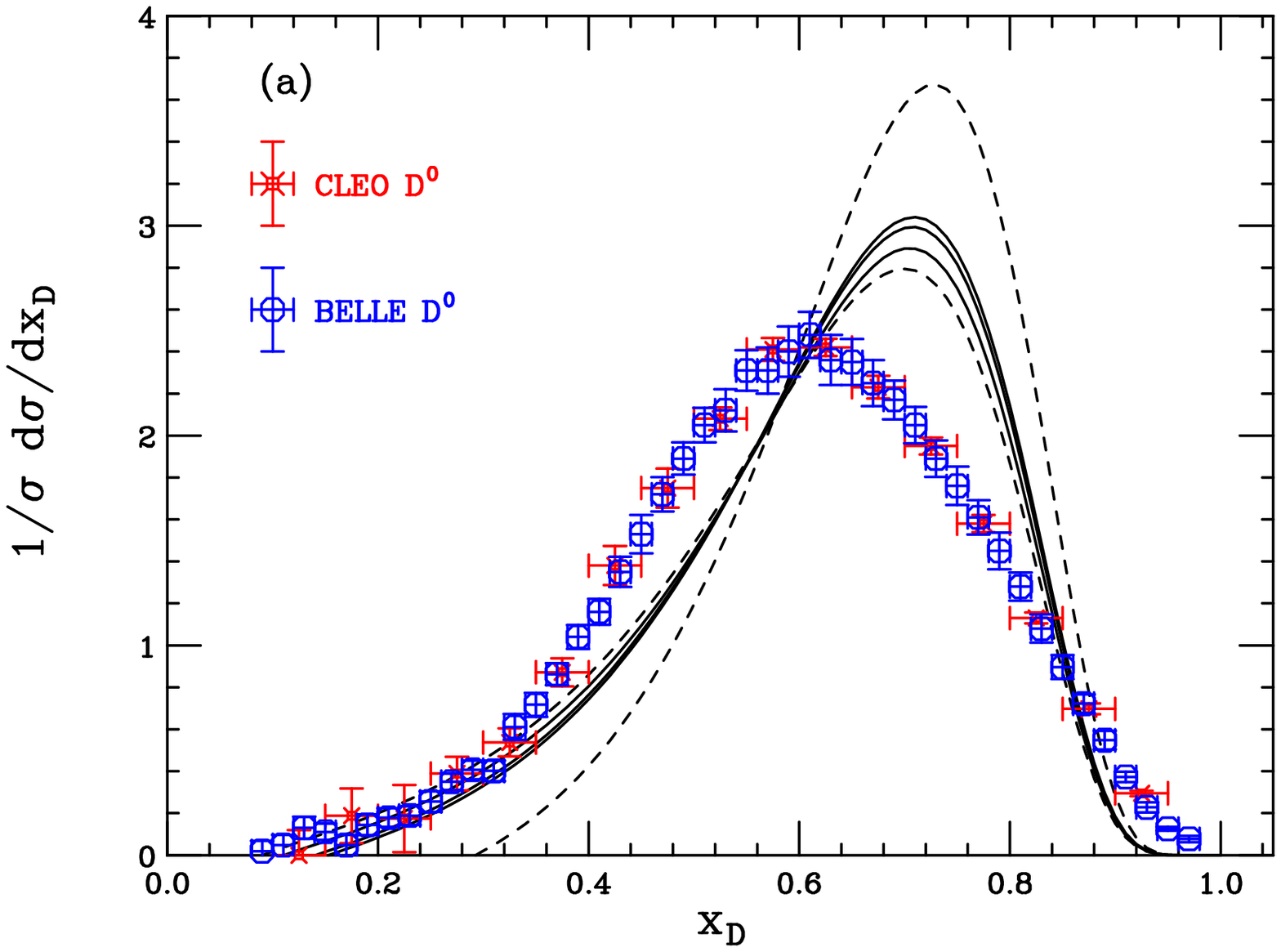,width=.49\textwidth}
    \epsfig{file=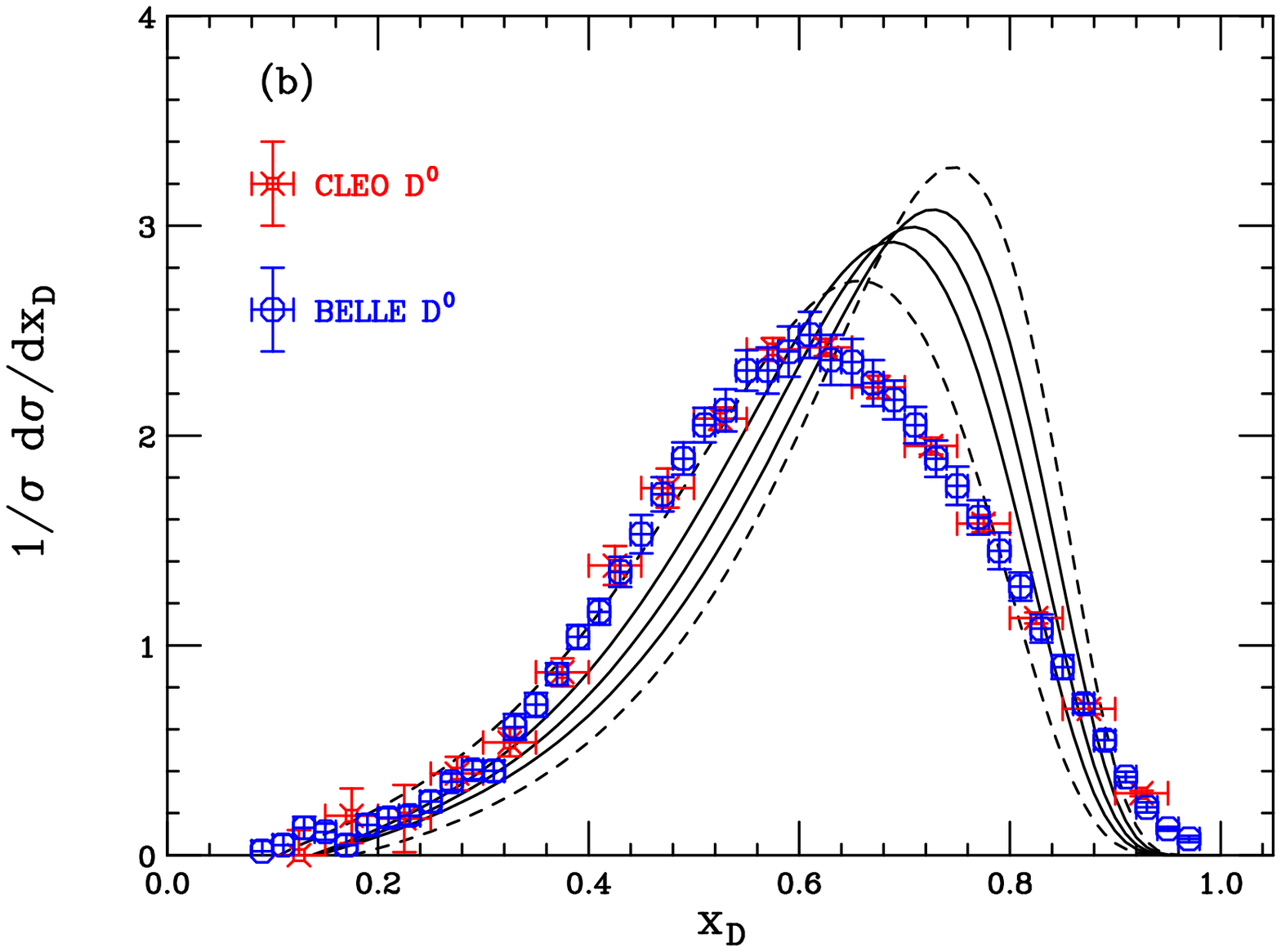,width=.49\textwidth}
\caption{As in figure \ref{aleph}, but comparing with
$D^0$ data from CLEO and BELLE experiments. 
In (a) we vary the factorization scales $\mu_F$ (solid) and
$\mu_{0F}$ (dashes); in (b) $\alpha_S(m_Z^2)$ (solid) and $m_c$ (dashes).}
\label{db}}


We present 
in figure \ref{dsb} the comparison of the predictions yielded
by our model with CLEO and BELLE data
on ${D^*}^0$-meson production. 
The comparison with the CLEO ${D^*}^0$ data, which
are affected by pretty small errors, is quite unsatisfactory
and the $\chi^2$ values high.
The BELLE spectrum instead exhibits larger errors,
so that we are able to compare with the data at $x_D<0.85$ with quite small
$\chi^2$ values. With our default parametrization,
we obtain $\chi^2/\mathrm{dof}=45.23/36$, while an even lower result,
$\chi^2/\mathrm{dof}=32.10/36$, is obtained if we set 
$\mu_{0F}=2m_c$, the same choice leading to the
best fit to the ALEPH ${D^*}^+$ spectrum.

We finally show in figure \ref{dbfp} the comparison of our predictions
with the data on ${D^*}^+$ production at CLEO and BELLE. Following
\cite{cno}, as far as the BELLE data are concerned, we present separately
the spectra of the mesons decaying according to ${D^*}^+\to D^+$ 
and ${D^*}^+\to D^0$, with the former presenting larger errors.
The comparison is qualitatively similar to figure \ref{dsb}, with
our model capable of describing well 
the data up to $x_D\simeq 0.6$, but failing
to reproduce the peak and the large-$x_D$ tail. 
Drawing a parallel between  figure \ref{dbfp} and
figure \ref{aleph}, where our model,
though within the uncertainties, led to a better comparison with respect to
the ALEPH ${D^*}^+$ data,  
one may argue that some major problems with
our approach seem to appear once the process hard scale decreases.
Later on, in section 6, we shall comment more about possible extensions of our
calculation and non-perturbative model, which may eventually improve the
comparison with the $B$-factory data in $x$-space.
\FIGURE{
    \epsfig{file=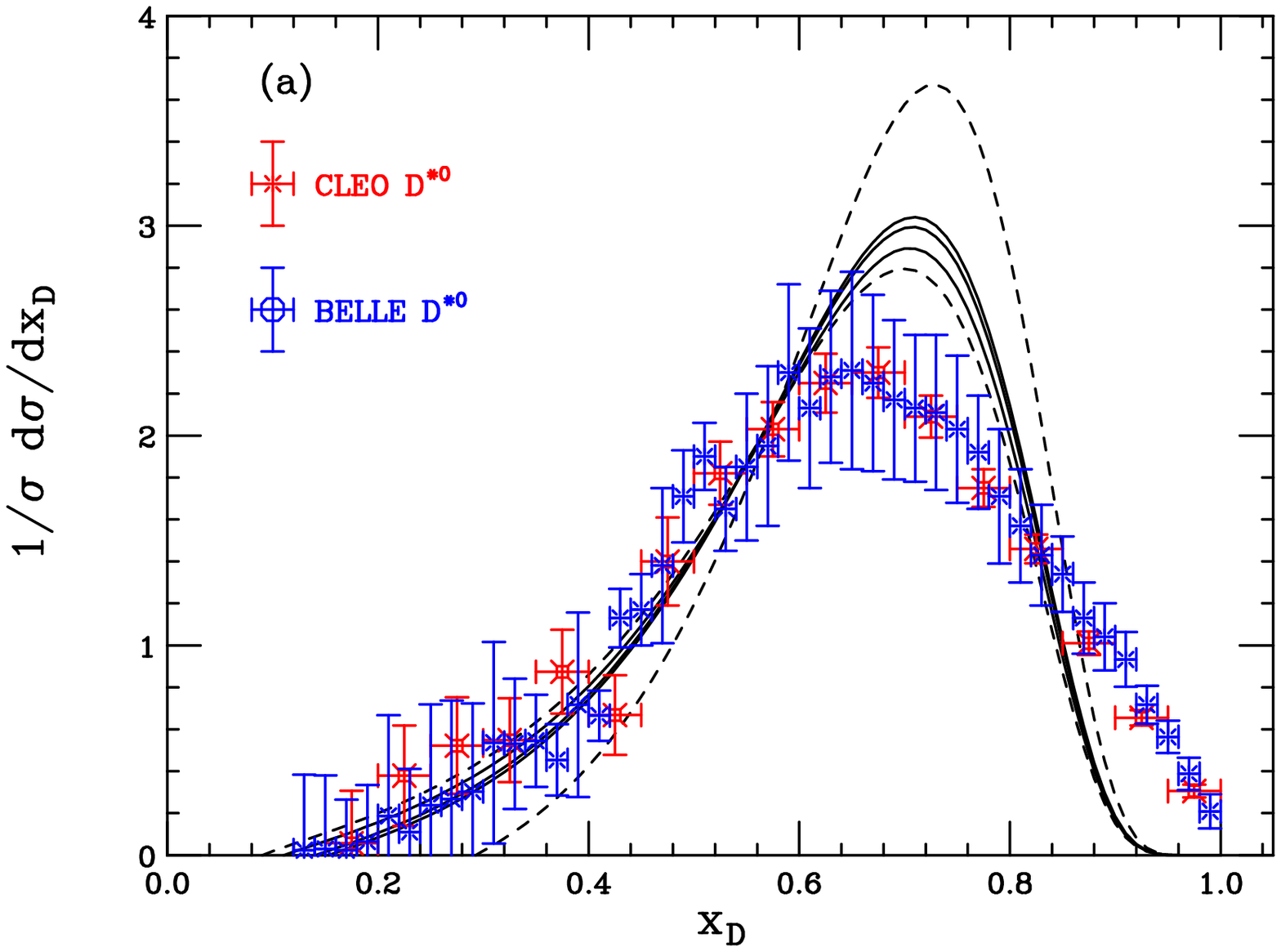,width=.49\textwidth}
    \epsfig{file=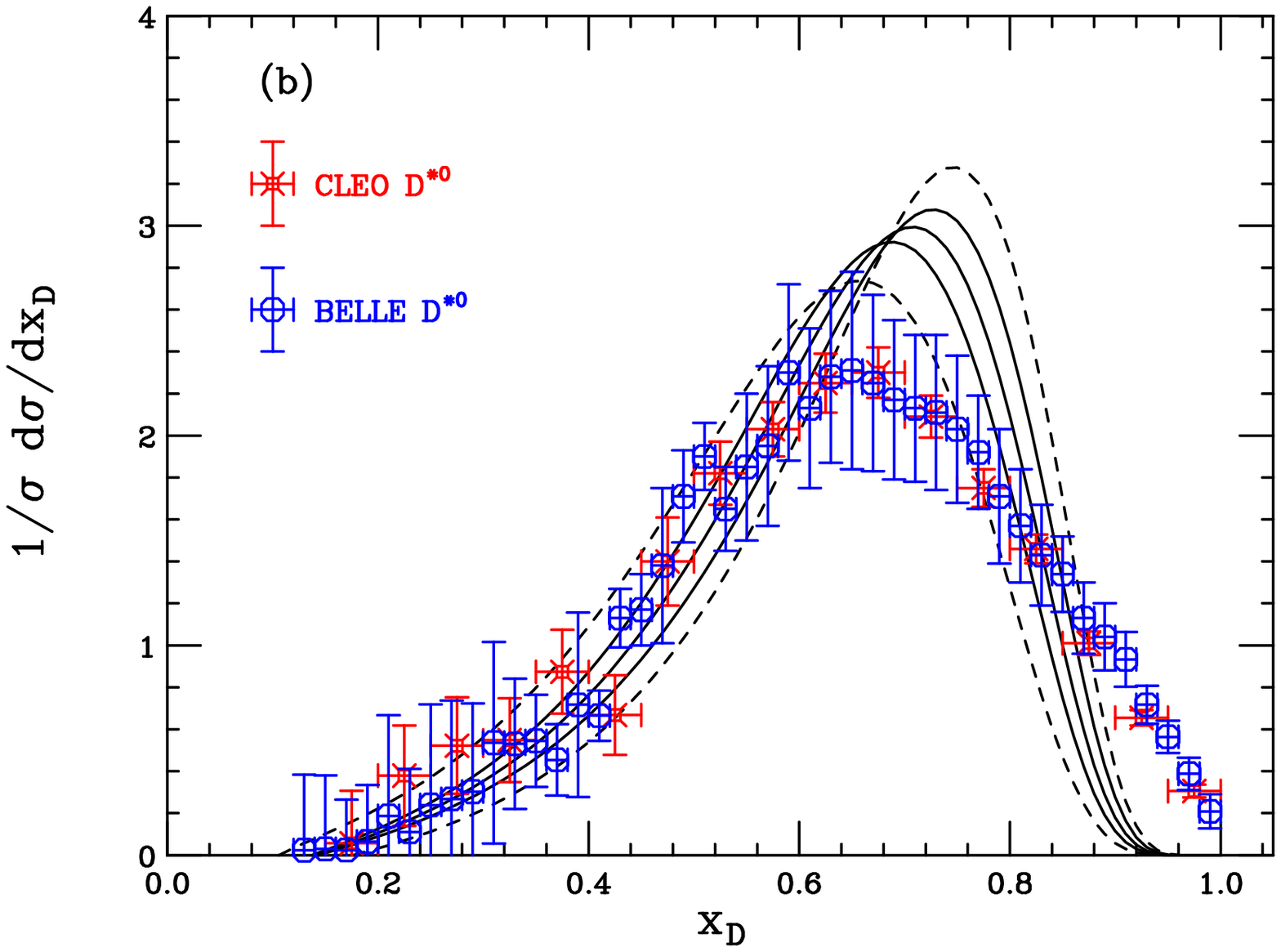,width=.49\textwidth}
\caption{Comparison of the prediction yielded by our
effective-coupling model with ${D^*}^0$ data from
BELLE and CLEO. The scales are varied as in figures \ref{aleph} and
\ref{db}.
\label{dsb}}}
\FIGURE{
    \epsfig{file=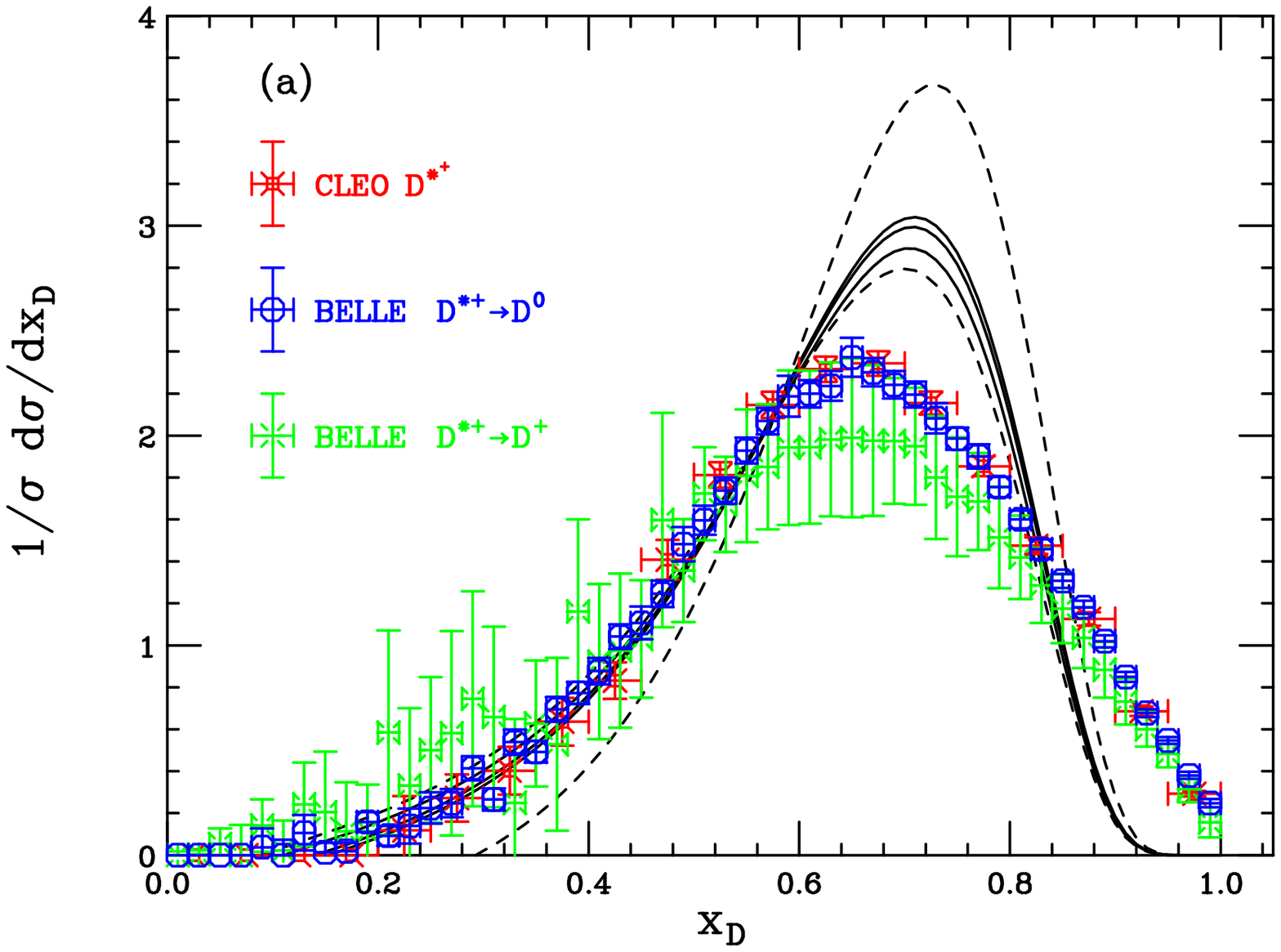,width=.49\textwidth}
    \epsfig{file=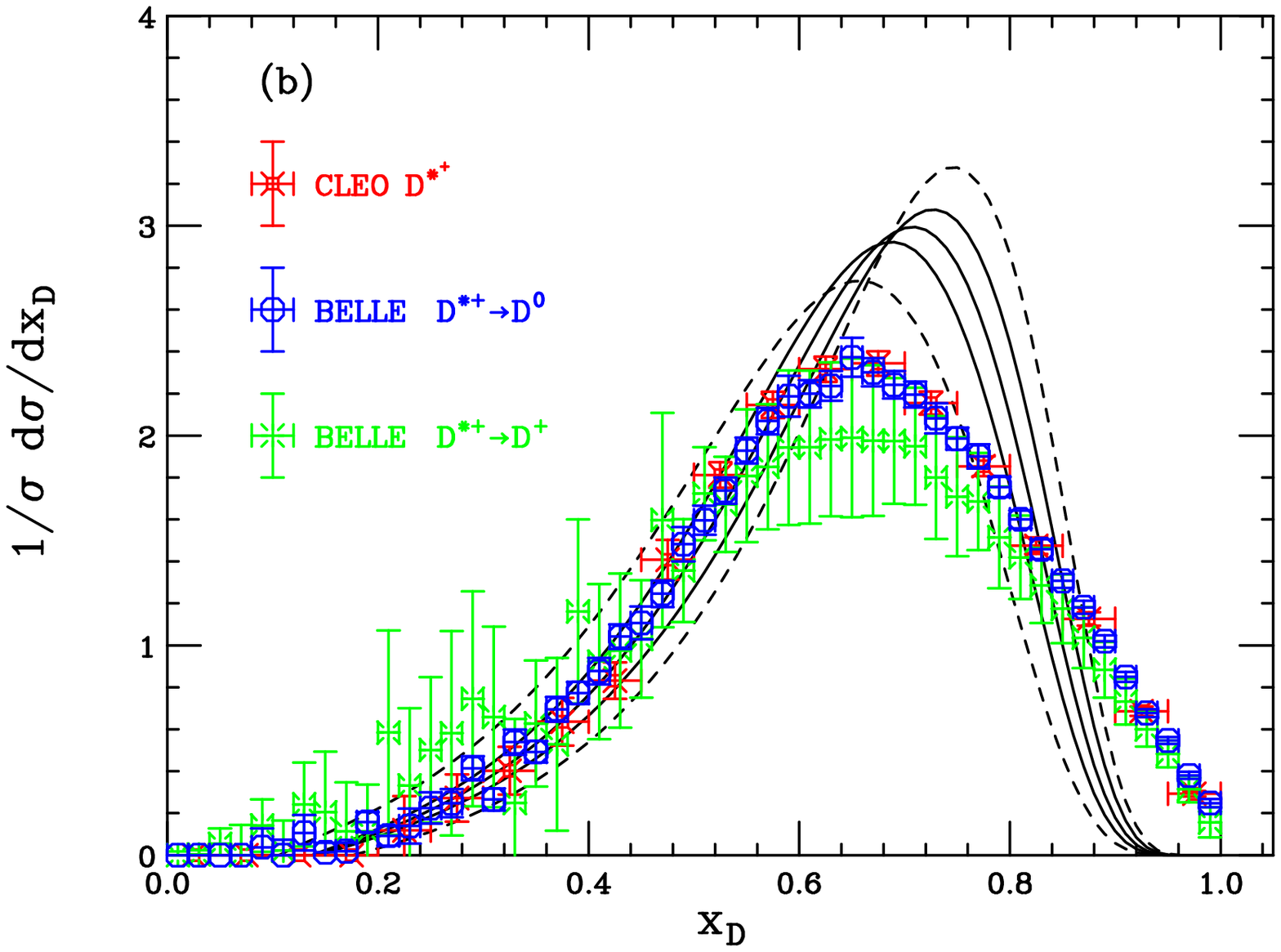,width=.49\textwidth}
\caption{As in figures \ref{db}--\ref{dsb}, but comparing 
with the CLEO and BELLE data on
${D^*}^+$ production. When presenting the BELLE data, we distinguish
the ${D^*}^+\to D^0$ from the ${D^*}^+\to D^0$ decay mode.
\label{dbfp}}}
\FIGURE{
 \label{pnp1}
\epsfig{file=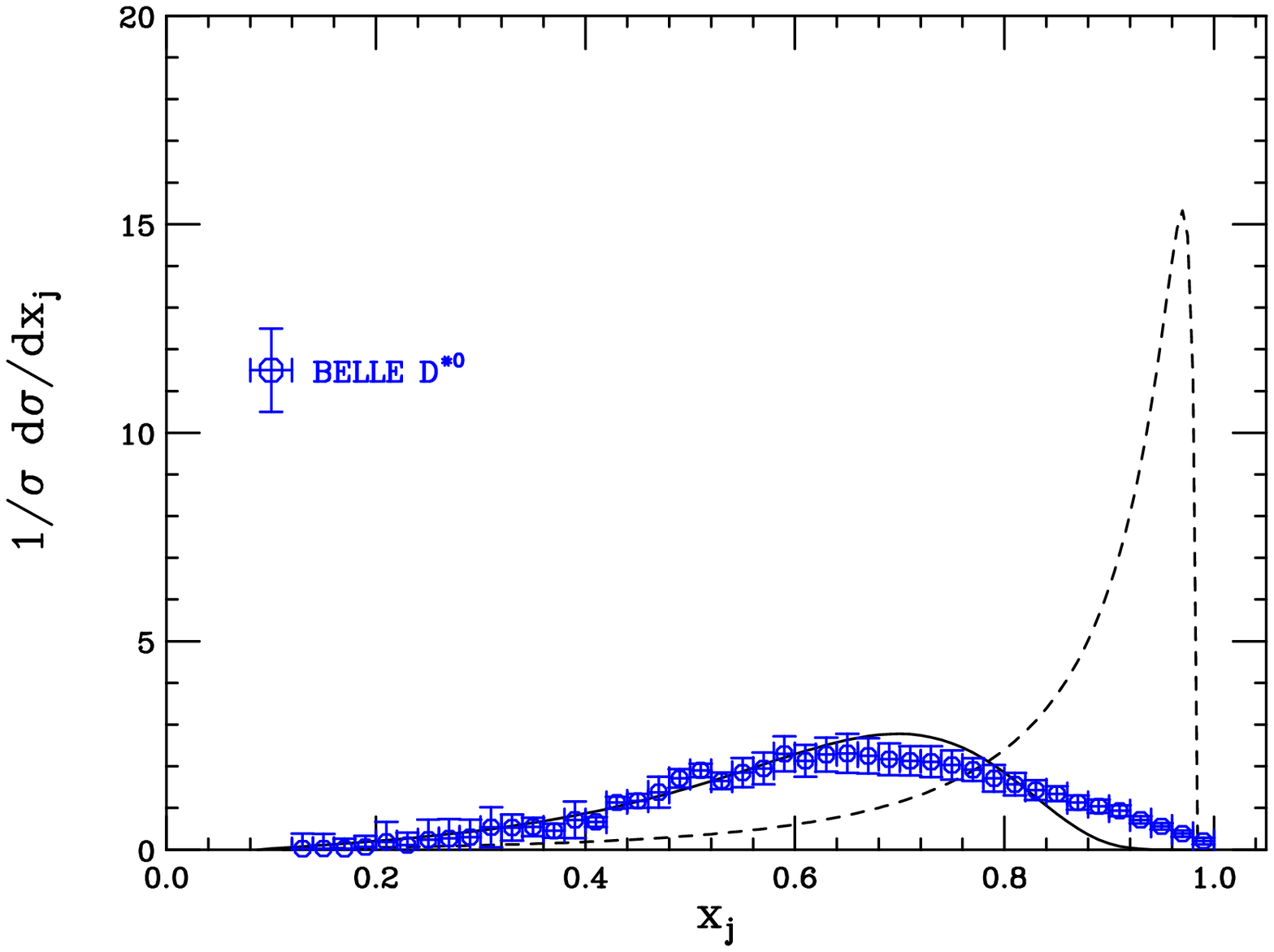, width=.7\textwidth}
\caption{The solid line is our best prediction
for charmed-hadron production at the $\Upsilon(4S)$ resonance ($j=D$);
the dashed line is the purely perturbative $c$-quark 
spectrum ($j=c$) yielded by the computation in \cite{cc}.
Also presented are the BELLE data on the ${D^*}^0$ spectrum.}}

As done when comparing with ALEPH,
we show in figure \ref{pnp1} the prediction leading to the
best fit to
the $B$-factory data, i.e. the one obtained for $\mu_{0F}=2m_c$,
along with the BELLE ${D^*}^0$ spectrum and the NLO/NLL perturbative
prediction from ref.~\cite{cc}.
We note that the parton-level result is sharply peaked at large $x$,
even more than in figure \ref{pnp2}: in fact, the smaller phase space available
at the $\Upsilon(4S)$ resonance with respect to the $Z^0$ pole
enhances the probability of producing $c\bar c$ pairs near the
threshold $x=1$.
Overall, the impact of non-perturbative corrections in the
coupling constant at the $\Upsilon(4S)$
resonance looks even more important than at LEP energies.

\sect{Results in $N$-space}

In this section we present our analysis in Mellin space and compare our
results with the experimental moments of the $D$-hadron cross section,
measured by ALEPH, CLEO and BELLE.
The data which we consider are the same as the ones which were
analysed in the previous section in $x$-space.
It was advocated in refs.~\cite{gardi,cn} that working in $N$-space can be 
theoretically preferable, as one does not need any explicit form for the 
non-perturbative
fragmentation function and its moments can be fitted directly from
the data. Of course, this issue does not apply to our case, since 
we are not fitting any parameter, but nonetheless it
is still worthwhile to compare our results 
with the experimental moments.
The $N$-space investigation will be particularly interesting at the 
$\Upsilon(4S)$ resonance, 
where the $x$-space
analysis has exhibited quite serious discrepancies.

\FIGURE{
\epsfig{file=alds.ps, width=.7\textwidth}
\caption{Moments of charmed-hadron cross section according to 
our effective-coupling model (denoted by `Theory'), 
compared with the moments of
${D^*}^+$ production at ALEPH. The theoretical errors are estimated by varying
the parameters entering in the perturbative calculation, 
as discussed throughout the text.}\label{mom1}}
\FIGURE{
\epsfig{file=bfd0.ps, width=.7\textwidth}
\caption{As in figure \ref{mom1}, but comparing our prediction with the
moments of the $D^0$ production cross section,
measured by the BELLE and CLEO collaborations.}
\label{mom2}}
\FIGURE{
\epsfig{file=bfds.ps, width=.7\textwidth}
\caption{As in figures \ref{mom1} and \ref{mom2},
with our results faring against the $N$-space 
${D^*}^0$ data from CLEO and BELLE.}\label{mom3}}
\FIGURE{
\epsfig{file=bfdsp.ps, width=.7\textwidth}
\caption{Comparison of our theoretical prediction
with the $N$-space
${D^*}^+$ data from CLEO and BELLE. As in the $x$-space analysis,
we plot separately the ${D^*}^+$ moments at BELLE, 
according to whether they decay via ${D^*}^+\to D^0$ or
${D^*}^+\to D^+$.}\label{mom4}}

We calculate the moments of the 
$D$-hadron cross section directly from the $N$-space formulas, i.e.
eq.~(\ref{sigmabhad}), and 
vary masses and scales as $x$-space, for the sake of
estimating the theoretical uncertainty.
The experimental moments are the same as the ones presented in
ref.~\cite{cno}, with the effect of electromagnetic 
initial-state radiation subtracted off. We just rescale them so that the
first moment of all data sets reads $\sigma_{N=1}=1$, as happens for
our theoretical results.
Following 
\cite{bfrag}, we first evaluate the uncertainties on the moments due to
to the variation of 
$\mu_F$, $\mu_{0F}$, $m_c$ and $\alpha_S(m_Z^2)$ separately,
and then estimate the overall theoretical error
summing in quadrature all individual uncertainties (see table~2 in
ref.~\cite{bfrag}).

The results of the comparison with the experimental moments are 
finally presented in
figures \ref{mom1}--\ref{mom4}, where 
we investigate how the prediction yielded 
by our model fares with respect to the moments of ALEPH ${D^*}^+$ 
(figure \ref{mom1}),  CLEO and BELLE $D^0$ (figure \ref{mom2}),
${D^*}^0$ (figure \ref{mom3}) and ${D^*}^+$ (figure \ref{mom4}) data.
Since our model does not distinguish spin and electric charge, we shall
always have the same theoretical moments, regardless of the kind of mesons
we are comparing with.
As found out in the $x$-space analysis, our predictions are affected by 
fairly large uncertainties; it is nonetheless 
interesting that, within the errors, the moments obtained using
the resummed calculation provided with the effective-coupling
model are compatible with the experimental ones. This result is
especially remarkable for the comparison with the data collected at the
$\Upsilon(4S)$ resonance (see figures \ref{mom2}--\ref{mom4}),
which exhibited instead relevant discrepancies in $x$-space. 
In fact, considering, e.g., the $x$-space $D^0$ spectra in figure \ref{db},
our model tends to underestimate
the event fraction at small and very large $x_D$ and overestimate
the differential cross section for $0.6\lsim x_D \lsim 0.8$.
Therefore, when evaluating integrated quantities like the moments,
such effects get compensated and one is able to obtain a reasonable description
of all $N$-space data (figure \ref{db}).
A similar result was indeed found in \cite{volker}, where
parton showers and resummed calculations 
were used to describe $B$-hadron production at the $Z^0$ pole.
The spectrum yielded by the
HERWIG Monte Carlo generator \cite{herwig} gave rise to
a quite large $\chi^2/{\rm dof}$ when comparing with $b$-fragmentation
data in $x$-space, even after tuning a few parameters.
However, HERWIG was able to reproduce fairly well the first few 
experimental moments of the $B$ cross section.

Referring, e.g., to the average value $\langle x_D\rangle$ of the
$D^0$ spectra at $B$-factories, corresponding to 
$\sigma_{N=2}$, the experimental data yield 
$\langle x_D\rangle=0.610\pm 0.005$ (CLEO) and $0.612\pm 0.006$ (BELLE).
Using our default perturbative parametrization and effective-coupling
model, we obtain instead $\langle x_D\rangle_{\rm th}=0.647$, above the 
experimental result.
Different choices of the parameters lead nevertheless to even larger or
lower estimates of $\sigma_{N=2}$. For example, setting $m_c=2.1$~GeV and
the other parameters to their default values, 
we obtain $\langle x_D\rangle_{\rm th}=0.686$, while for $m_c=1.5$~GeV
the second moment reads $\langle x_D\rangle_{\rm th}=0.599$, below the
CLEO and BELLE data.
In any case, as already
pointed out for the purpose of the $x$-space analysis, a 
complete NNLO/NNLL should possibly 
decrease the theoretical error in Mellin space as well.
Moreover, since even the best fits in $x$-space were obtained discarding
the data points at $x_D>1-\Lambda/m_c$, we expect that the comparison 
with the experimental moments should eventually
get worse for very large values of $N$, dominated by the $x_D\gsim 0.85$ region
in $x$-space, where our
predictions are systematically below the experimental data.
A hint for  such a behaviour can be learned from
figures \ref{mom3} and \ref{mom4}, where already
the $\sigma_{N=10}$ theoretical moment is
only marginally consistent with the data, even
within the uncertainties. 

\sect{Conclusions}
We studied charm-quark fragmentation in $e^+e^-$ annihilation and used
a recently proposed model, based on an effective strong coupling constant,
as the only source of non-perturbative effects.
Such a model was already employed in \cite{afr,bfrag} and gave a reasonable
description of $b$-quark fragmentation in $e^+e^-$ annihilation
and some $B$-meson decay data.
We described charm-quark perturbative production following
the perturbative fragmentation approach, with NLO coefficient function and
initial condition of the perturbative fragmentation function, NLL
DGLAP non-singlet evolution and NNLL large-$x$ resummation.
Resummed expressions were matched to the exact NLO ones using
the so-called $\ln R$-prescription, which turned out
to significantly improve the spectrum near the $x=1$ endpoint. 
The effective coupling was implemented in the NNLO approximation,
as in refs.~\cite{afr,bfrag}.

We compared the predictions of our model with data from ALEPH, BELLE and
CLEO, corrected for initial-state photon-radiation effects as in
\cite{cno}. Throughout our analysis, 
since our non-perturbative model has no tunable
parameter, we varied the quantities in the perturbative calculation within
typical ranges, according to the values quoted in \cite{pdg}. 

We found that our model is able to acceptably describe, 
for $x_D<1-\Lambda/m_c$ and within the
theoretical and experimental errors, the  
${D^*}^+$ spectrum from ALEPH.
In particular, the best fits to the data are obtained, within
our chosen ranges, if we set the factorization scale entering in 
the initial condition to $\mu_{0F}=2m_c$. A value of $m_c$ consistent
with the charm pole mass, rather than the $D$-meson mass, improves
the comparison at small $x_D$ and around the peak.
Significant discrepancies were instead found with respect to the
$D^0$, ${D^*}^0$ and ${D^*}^+$
data from $B$-factories, where we succeeded in
obtaining $\chi^2/{\rm dof}\simeq 1$ only when comparing with the
BELLE ${D^*}^0$ spectrum, affected by pretty large errors.
The experimental data on $D^0$ and ${D^*}^+$ production
at CLEO and BELLE and on ${D^*}^0$ at CLEO
exhibit instead very small errors and we did not manage to obtain a reasonable 
$\chi^2/\mathrm{dof}$, even within the theory error.
We just noticed that setting $m_c=1.5$~GeV 
gives a good description of the data for $x_D<0.6$, but some major
disagreement is still present for larger values of $x_D$.
In Mellin space, however, within the fairly large theoretical
uncertainties, we managed to reproduce the first ten moments of all
considered data samples. 
We expect nonetheless that for larger values of
$N$ some discrepancy should appear, consistently with the observation that
even the best $x$-space fits were obtained discarding few
large-$x_D$ data points.

Anyway, we remind that some problems with reproducing both
ALEPH and $B$-factory data were already encountered in ref.~\cite{cno},
where the authors employed a NLO/NLL calculation,
a non-perturbative fragmentation
function with three free parameters and rescaled $N$ according to 
eq.~(\ref{nprime}).
The hadronization model was tuned to reproduce fairly
well all $B$-factory data on neutral as well as charged $D$- and
$D^*$-meson production. Nevertheless, the best-fit parametrization
did not succeed in reproducing the ALEPH ${D^*}^+$ spectrum after evolving to
LEP energies. The conclusion of the analysis carried out in \cite{cno}
was that, in order to reconcile both LEP and $B$-factory data, it was
necessary to include power corrections in the process-dependent 
coefficient function, depending on the process hard scale.
This way, one should be able to describe all data, still using the same 
perturbative accuracy and the same functional form for the non-perturbative 
part. 
However,  
due to the errors in the intermediated $x_D$ region, ref.~\cite{cno}
was not able to discriminate whether the missing power corrections should 
behave according to a $1/Q$ or a $1/Q^2$ power law, with $Q$ being the
centre-of mass energy. In any case, being $m_Z$ much larger than 
$m_{\Upsilon(4S)}$, one should expect that such a power correction should 
mainly modify the spectra at the $\Upsilon(4S)$ resonance, 
in such a way that the fits 
of the non-perturbative fragmentation functions at the $Z^0$ pole,
presented in \cite{cno}, should eventually work even at 
$B$-factory energies, after minimal adjustments \cite{matteo}.


As far as our work is concerned, we do find it interesting that, although
within the theoretical and experimental errors and 
after discarding few data points at very large $x_D$, our 
parameter-free model yields  $\chi^2/\mathrm{dof}\simeq 1$
from the comparison with ALEPH ${D^*}^+$ data 
and reproduces the moments 
of all analysed data sets.
The  discrepancies of our prediction
with respect to the very precise data from 
CLEO and BELLE in $x$-space clearly deserve further investigation. 
The results in this paper, along with the ones reported in \cite{bfrag}, 
seem to indicate that the model works better for
heavy-quark fragmentation 
at the $Z^0$ pole, while more serious
problems show up once the hard scale is lowered.
However, only a power correction, such as the one understood in \cite{cno},
mostly relevant at large $x$ or $N$, may not to be 
enough to solve
the discrepancy with the $B$-factory data, as  
figures \ref{db}--\ref{dbfp} 
show disagreement even around the peak and at small $x_D$.


The theoretical uncertainty is expected to decrease 
after the inclusion of NNLO coefficient functions \cite{neerven,moch}, 
initial condition \cite{alex1,alex3} and non-singlet splitting functions
\cite{alex2}, which will also promote DGLAP evolution to NNLL
accuracy in the non-singlet sector.
Since within our approach  we are including power corrections in an
effective coupling, any perturbative improvement, such as accounting for
${\cal O}(\tilde\alpha_S^2)$ contributions, will necessarily imply the
inclusion of non-perturbative corrections as well.
Furthermore, 
the change (\ref{a3}) in the coefficient $A^{(3)}$ 
has been 
implemented in the threshold NNLL expressions, but not yet
in the splitting functions, whose 
NNLO corrections do contain a contribution $\sim A^{(3)}$ \cite{alex2}.
Including such a term in the splitting functions,
along with the redefinition  $A^{(3)}\to \tilde A^{(3)}$,
may shift the $x_D$ spectrum and possibly improve the comparison 
with the $B$-factory data, as found in \cite{bfrag} for 
$B$-hadron production at LEP and SLD.
We should also expect a relevant impact on our analysis of the possible
inclusion of large-$x$ next-to-next-to-next-to-leading logarithmic
(NNNLL) terms, whose coefficients have been denoted by
$A^{(4)}$, $B^{(3)}$
and $D^{(3)}$ in Eqs.~(\ref{GNC}) and (\ref{resdini}).
All such coefficients will be modified in a fashion
analogous to eq.~(\ref{a3}) when using $\tilde \alpha_S(k^2)$.
Moreover, the implementation of higher-order 
threshold contributions in the exponents (\ref{GNC}) and (\ref{resdini})
will also lead to the inclusion of  further
power corrections since, as part of our model, we performed the Mellin
transforms exactly.



The other guideline to obtain better agreement with the
$x$-space data consists in modifying the effective-coupling model, e.g.,
introducing a correcting term $\delta\tilde\alpha_S(k^2)$ 
as in eq.~(\ref{dal}),
possibly containing extra parameters.
In order to accommodate both $D$ and $D^*$ data,  $\delta\tilde\alpha_S(k^2)$
may possibly depend on the spin of the considered hadron.
However, 
before speculating about its functional form, we believe that
we still need a NNLO/NNLL calculation
to reduce the theoretical uncertainty and deal with a more
stable prediction. In fact, without a NNLO/NNLL analysis, 
function 
$\delta\tilde\alpha_S(k^2)$ will largely depend on the values chosen for
the perturbative parameters and considerably vary according to
whether, e.g., one sets $m_c=1.5$ or 2.1 GeV, $\mu_{0F}=m_c/2$
or $2m_c$, and so on.

The large-$x_D$ behaviour of our spectra may  be improved as well, since even
the best comparisons with the data were obtained 
in this paper for $x_D<1-\Lambda/m_c$ and in \cite{bfrag} for
$x_B<1-\Lambda/m_b$.
An option could be 
the prescription (\ref{nprime}) suggested in \cite{cno}; in fact,
any modification at large $x_D$ will indirectly affect, via normalization, the 
energy distribution at smaller $x_D$ as well. 
Nevertheless, once again, given the 
uncertainties exhibited by our predictions even at 
large $x_D$, this investigation should be better performed using
a calculation of higher accuracy.


Ideally, once the above issues are clarified,
one may think of using 
our model to describe $D$- and $B$-hadron 
production at the Tevatron accelerator, along the lines of 
refs.~\cite{cn,frix}, and extend the results to LHC energies.
Nevertheless, unlike the standard analyses, where a non-perturbative
fragmentation function is fitted to $e^+e^-$ data and then used in the
hadron-collider environment, we are not tuning any parameter to 
the $e^+e^-$ data.
Therefore, possible 
studies at hadron colliders will be 
independent checks of the capability of
our model to reproduce heavy-quark fragmentation data.

Moreover, we can use the NLO perturbative calculations
in \cite{ccm,corc}, along with the 
effective coupling constant, to predict bottomed-hadron 
spectra in top ($t\to bW$) or 
Higgs ($H\to b\bar b$) decays at the Tevatron and LHC.
Finally, the $c$-fragmentation
results here presented can be compared with the ones
yielded by Monte Carlo generators, extending the
analysis carried out in \cite{volker}, where parton shower algorithms and 
resummations were compared for the purpose of $B$-hadron production
in $e^+e^-$ annihilation, top and Higgs decays.
For such a comparison to be
consistent, however, even the HERWIG \cite{herwig} and PYTHIA \cite{pythia}
generators will have to be tuned to the same 
LEP and $B$-factory data analysed throughout this paper.
It will also be very interesting to
implement the effective coupling constant to replace, e.g., the cluster
model \cite{bryan} 
which simulates the hadronization in HERWIG and investigate how the 
Monte Carlo results fare with respect to the experimental data
on $D$- and $B$-hadron production.
This is in progress as well.

\paragraph{Acknowledgements.}

\noindent
We are indebted to U.~Aglietti for a series of very useful discussions on
the effective-coupling model.
We acknowledge D.~De Florian for discussions on resummed
calculations and
M.~Cacciari for many conversations on the perturbative
fragmentation approach and for providing us with the computing
code to obtain the results of ref.~\cite{cc} presented in figures \ref{pnp2}
and \ref{pnp1}.
We thank C.~Oleari for proving us with the data presented in
ref.~\cite{cno}, accounting for initial-state radiation effects. 
This work was partially supported by ALFA-EC funds in the framework
of Program HELEN (High Energy Physics Latinoamerican-European Network).
G.F. acknowledges support by the European Community's Marie-Curie
Research Training Network Programme under contract MRTN-CT-2006-035505
``Tools and Precision Calculations for Physics Discoveries at Colliders''
G.F. is also grateful to the Physics Department of the University of Buenos
Aires for warm hospitality during some of this work.


\begin{thebibliography}{99}

\bibitem{kart}
V.G. Kartvelishvili, A.K. Likehoded and V.A. Petrov, 
\emph{On the fragmentation functions of heavy quarks into hadrons},
\plb{78}{1978}{615}.

\bibitem{pet}
C. Peterson, D. Schlatter, I. Schmitt and P.M. Zerwas, 
\emph{Scaling violations in inclusive $e^+e^-$ annihilation spectra},
\prd{27}{1983}{105}.


\bibitem{afr}
U. Aglietti, G. Ferrera and G. Ricciardi,  
\emph{Semi-inclusive B decays and a model for soft-gluon effects},
\npb{768}{2007}{85}.

\bibitem{bfrag}
U. Aglietti, G. Corcella and G. Ferrera, 
\emph{Modelling non-perturbative corrections to bottom-quark fragmentation},
\npb{775}{2007}{162}.

\bibitem{shirkov}
D. Shirkov,  \emph{Nonpower expansions for QCD observables at low energies},
\npps{152}{2006}{51}.



\bibitem{stefanis}
  N.~G.~Stefanis, W.~Schroers and H.~C.~Kim,  
\emph{Pion form-factors with improved infrared factorization},
\plb{449}{1999}{299}; 


 N.~G.~Stefanis, W.~Schroers and H.~C.~Kim,
\emph{Analytic coupling and Sudakov effects in exclusive processes: 
pion and $\gamma^*\gamma\to\pi^0$ form factors},
\epjc{18}{2000}{137}.


\bibitem{mele}
B. Mele and P. Nason, \emph{The fragmentation function for heavy quarks in 
QCD}, \npb{361}{1991}{626}.

\bibitem{marti}
G. Altarelli, R.K. Ellis, G. Martinelli and S.--Y. Pi, 
\emph{ 
Processes involving fragmentation functions beyond the leading order in QCD},
\npb{160}{1979}{301}.

\bibitem{dgl}
V.N. Gribov and L.N. Lipatov,  
\emph{Deep inelastic $ep$ scattering in perturbation theory},
\sjnp{15}{1972}{438};


L.N. Lipatov, \emph{The parton model and perturbation theory},
\sjnp{20}{1975}{95};


Yu.L. Dokshitzer, \emph{Calculation of the structure functions for
deep inelastic scattering and $e^+e^-$ annihilation by perturbation
theory in quantum chromodynamics (in Russian)},
\jetp{46}{1977}{298}. 

\bibitem{ap}
G. Altarelli and G. Parisi,  
\emph{Asymptotic freedom in parton language},
\npb{126}{1977}{298}.


\bibitem{cc} 
M. Cacciari and S. Catani, 
\emph{Soft-gluon resummation 
for the fragmentation of light and heavy quarks at large $x$},
\npb{617}{2001}{253}.


\bibitem{neerven}
P.J. Rijken and W.L. van Neerven,  
\emph{Higher-order QCD corrections to the transverse and longitudinal 
fragmentation functions in electron-positron annihilation},
\npb{487}{1997}{233}.

\bibitem{moch}
A.D. Mitov and S.O. Moch,
\emph{QCD corrections to semi-inclusive hadron production in electron positron
annihilation at two loops}, \npb{751}{2006}{18}.


\bibitem{alex1}
K. Melnikov and A.D. Mitov, 
\emph{Perturbative heavy quark fragmentation function through 
${\cal O}(\alpha_S^2)$}, \prd{70}{2004}{034027}.


\bibitem{alex2}
A.D. Mitov, S. Moch and A. Vogt, 
\emph{Next-to-next-to-leading order evolution of non-singlet fragmentation
functions},
\plb{638}{2006}{61}.

\bibitem{alex3} 
A.D. Mitov, \emph{Perturbative heavy quark fragmentation function 
through ${\cal O}(\alpha_S^2)$: Gluon initiated contribution},
\prd{71}{2005}{054021}.


\bibitem{sterman}
G. Sterman,  
\emph{Summation of large corrections to short distance hadronic cross-
sections}, \npb{281}{1987}{310}.


\bibitem{ct}
S. Catani and L. Trentadue, \emph{Resummation of the QCD perturbative 
series for hard processes}, \npb{327}{1989}{323}.


\bibitem{moch1}
S. Moch, J.A.M. Vermaseren and A. Vogt, 
\emph{The three loop splitting functions in QCD: the nonsinglet case},
\npb{688}{2004}{101}.

\bibitem{moch2}
S. Moch, J.A.M. Vermaseren and A. Vogt, 
\emph{Nonsinglet structure functions at three loops: fermionic contributions},
\npb{646}{2002}{181}.

\bibitem{march}
G.P. Korchemsky and G. Marchesini,
\emph{Structure function for large x and renormalization of Wilson loop},
\npb{406}{1993}{225}.

\bibitem{gardi1}
 E. Gardi, \emph{On the quark distribution in an on-shell heavy quark and its 
all-order relations with the perturbative fragmentation function},
\jhep{0502}{2005}{053}.


\bibitem{logr}
 S. Catani, G. Turnock, B.R. Webber and L. Trentadue,  
\emph{Thrust distribution in $e^+e^-$ annihilation}, \plb{263}{1991}{491}.

\bibitem{gardi}
M. Cacciari and E. Gardi, \emph{Heavy quark fragmentation},
\npb{664}{2003}{299}.


\bibitem{deflo}
S. Catani, D. de Florian, M. Grazzini and P. Nason,  
\emph{Soft gluon resummation for Higgs boson production at hadron colliders},
\jhep{07}{2003}{028}.

\bibitem{agl}
U. Aglietti and G. Ricciardi, 
\emph{A model for next-to-leading order threshold resummed 
form-factors}, \prd{70}{2004}{114008}.

\bibitem{braun}
M. Beneke and V.M. Braun,  
\emph{Power corrections and renormalons in Drell--Yan production}, 
\npb{454}{1995}{253}.

\bibitem{amati} 
D. Amati, A. Bassetto, M. Ciafaloni, G. Marchesini and
G. Veneziano,  
\emph{A treatment of hard processes sensitive to the infrared structure of 
QCD}, \npb{173}{1980}{429}.

\bibitem{min}
S. Catani, M.L. Mangano, P. Nason and L. Trentadue,  
\emph{The resummation of soft gluons in hadronic collisions},
\npb{478}{1996}{273}.

\bibitem{cno} 
M. Cacciari, P. Nason and C. Oleari,  
\emph{A study of heavy flavoured meson fragmentation functions in 
$e^+ e^-$ annihilation}, \jhep{0604}{2006}{006}.



\bibitem{lep}
ALEPH collaboration, R. Barate et al.,  
\emph{Study of charm production in Z decays}, \epjc{16}{2000}{597}.

\bibitem{cleo}
CLEO collaboration, M.~Artuso et al., 
\emph{Charm meson spectra in $e^+e^-$ 
annihilation at 10.5~GeV c.m.e.}, \prd{70}{2004}{112001}.

\bibitem{belle}
BELLE collaboration, R. Seuster  et al., \emph{Charm hadrons from 
fragmentation and B decays in $e^+ e^-$ annihilation at $\sqrt{s} = 10.6$~GeV},
\prd{73}{2006}{032002}. 

\bibitem{pdg}
Particle Data Group collaboration, W.~M.~Yao et al.,  
\emph{Review of Particle Physics}, \jphg{33}{2006}{1}.

\bibitem{cno1}
M. Cacciari, P. Nason and C. Oleari,  
\emph{Crossing heavy-flavour thresholds in fragmentation function}, 
\jhep{0510}{2005}{034}.

\bibitem{volker}
G. Corcella and V. Drollinger, 
\emph{Bottom-quark fragmentation: Comparing 
results from tuned event  generators  and resummed calculations} 
\npb{730}{2005}{82}.

\bibitem{herwig}
G. Corcella, I.G. Knowles, 
G. Marchesini, S. Moretti, 
K. Odagiri, P. Richardson, M.H. Seymour, B.R. Webber,
\emph{HERWIG 6: An event generator for hadron emission reactions with
interfering gluons (including supersymmetric processes)},
\jhep{0101}{2001}{010}.

\bibitem{matteo}
M. Cacciari, private communication.

\bibitem{cn}
M. Cacciari and P. Nason,  
\emph{Charm cross sections for the Tevatron Run II}, \jhep{0309}{2003}{006}.


\bibitem{frix}
M. Cacciari, S. Frixione, M.~L. Mangano, P. Nason and G. Ridolfi,
\emph{QCD analysis of first $B$ cross-section data at 
1.96 TeV}, \jhep{0407}{2004}{033}.


\bibitem{ccm}
M. Cacciari, G. Corcella and A.D. Mitov,  
\emph{Soft-gluon resummation for bottom fragmentation in top quark decay},
\jhep{0212}{2002}{015}.


\bibitem{corc}
G. Corcella, \emph{Fragmentation in $H\to b\bar b$ processes}, 
\npb{705}{2005}{363}, Erratum  \ibid 
{713}{2005}{609}.


\bibitem{pythia}
T. Sjostrand, S. Mrenna and P. Skands,
\emph{PYTHIA 6.4 physics and manual}, \jhep{0605}{2006}{026}.

\bibitem{bryan}
B.R. Webber, \emph{A QCD model for 
jet fragmentation including soft gluon interference}, \npb{238}{1984}{492}.

\end{thebibliography}
\end{document}